\documentclass[%
reprint,
amsmath,amssymb,
aps,
pra,
superscriptaddress,
twocolumn,
tightenlines,
]{revtex4-2}

\usepackage{graphicx}% Include figure files
\usepackage{dcolumn}% Align table columns on decimal point
\usepackage{bm}% 
\usepackage{dcolumn}% Align table columns on decimal point
\usepackage{siunitx}
\usepackage{amsmath}
\usepackage{amssymb}
\usepackage{booktabs}
\usepackage{setspace}
\usepackage{xcolor}
\usepackage{lineno}
\begin{document}

\title{Flat band-engineered spin-density wave and the emergent multi-\textit{k} magnetic state in the topological kagome metal Mn$_{3}$Sn}

\author{Xiao Wang}
\thanks{These authors contributed equally to this work}
\affiliation{J\"ulich Centre for Neutron Science (JCNS) at Heinz Maier-Leibnitz Zentrum (MLZ), Forschungszentrum J\"ulich, Lichtenbergstrasse 1, D-85747 Garching, Germany}

\author{Fengfeng Zhu}
\thanks{These authors contributed equally to this work}
\affiliation{J\"ulich Centre for Neutron Science (JCNS) at Heinz Maier-Leibnitz Zentrum (MLZ), Forschungszentrum J\"ulich, Lichtenbergstrasse 1, D-85747 Garching, Germany}
\affiliation{State Key Laboratory of Functional Materials for Informatics, Shanghai Institute of Microsystem and Information Technology, Chinese Academy of Sciences, 200050 Shanghai, China}

\author{Xiuxian Yang}
\thanks{These authors contributed equally to this work}
\affiliation{Key Laboratory of Advanced Optoelectronic Quantum Architecture and Measurement (Ministry of Education), Beijing Key Laboratory of Nanophotonics and Ultrafine Optoelectronic Systems, and School of Physics, Beijing Institute of Technology, Beijing 100081, China}

\author{Martin Meven}
\affiliation{J\"ulich Centre for Neutron Science (JCNS) at Heinz Maier-Leibnitz Zentrum (MLZ), Forschungszentrum J\"ulich, Lichtenbergstrasse 1, D-85747 Garching, Germany}
\affiliation{Institut f\"ur Kristallographie, RWTH Aachen University, D-52056 Aachen, Germany}

\author{Xinrun Mi}
\affiliation{Low Temperature Physics Lab, College of Physics \& Center of Quantum Materials and Devices, Chongqing University, Chongqing 401331, China}

\author{Changjiang Yi}
\affiliation{Beijing National Laboratory for Condensed Matter Physics and Institute of Physics, Chinese Academy of Sciences, Beijing 100190, China}

\author{Junda Song}
\affiliation{J\"ulich Centre for Neutron Science (JCNS) at Heinz Maier-Leibnitz Zentrum (MLZ), Forschungszentrum J\"ulich, Lichtenbergstrasse 1, D-85747 Garching, Germany}

\author{Thomas Mueller}
\affiliation{J\"ulich Centre for Neutron Science (JCNS) at Heinz Maier-Leibnitz Zentrum (MLZ), Forschungszentrum J\"ulich, Lichtenbergstrasse 1, D-85747 Garching, Germany}

\author{Wolfgang Schmidt}
\affiliation{J\"ulich Centre for Neutron Science (JCNS) at ILL, Forschungszentrum J\"ulich, F-38000 Grenoble, France}

\author{Karin Schmalzl}
\affiliation{J\"ulich Centre for Neutron Science (JCNS) at ILL, Forschungszentrum J\"ulich, F-38000 Grenoble, France}

\author{Eric Ressouche}
\affiliation{Universit\'e Grenoble Alpes, CEA, IRIG, MEM, MDN, F-38000 Grenoble, France}

\author{Jianhui Xu}
\affiliation{Helmholtz-Zentrum Berlin für Materialien und Energie GmbH, Hahn-Meitner-Platz 1, D-14109 Berlin, Germany}

\author{Mingquan He}
\affiliation{Low Temperature Physics Lab, College of Physics \& Center of Quantum Materials and Devices, Chongqing University, Chongqing 401331, China}

\author{Youguo Shi}
\affiliation{Beijing National Laboratory for Condensed Matter Physics and Institute of Physics, Chinese Academy of Sciences, Beijing 100190, China}

\author{Wanxiang Feng}
\email{E-mail: wxfeng@bit.edu.cn}
\affiliation{Key Laboratory of Advanced Optoelectronic Quantum Architecture and Measurement (Ministry of Education), Beijing Key Laboratory of Nanophotonics and Ultrafine Optoelectronic Systems, and School of Physics, Beijing Institute of Technology, Beijing 100081, China}

\author{Yuriy Mokrousov}
\affiliation{Peter Gr\"unberg Institut (PGI) and Institute for Advanced Simulation (IAS), Forschungszentrum J\"ulich and JARA, D-52425 J\"ulich, Germany}
\affiliation{Institute of Physics, Johannes Gutenberg University Mainz, D-55099, Mainz, Germany}

\author{Stefan Bl\"ugel}
\affiliation{Peter Gr\"unberg Institut (PGI) and Institute for Advanced Simulation (IAS), Forschungszentrum J\"ulich and JARA, D-52425 J\"ulich, Germany} 

\author{Georg Roth}
\affiliation{Institut f\"ur Kristallographie, RWTH Aachen University, D-52056 Aachen, Germany}

\author{Yixi Su}
\email{E-mail: y.su@fz-juelich.de}
\affiliation{J\"ulich Centre for Neutron Science (JCNS) at Heinz Maier-Leibnitz Zentrum (MLZ), Forschungszentrum J\"ulich, Lichtenbergstrasse 1, D-85747 Garching, Germany}

\date{\today}% It is always \today, today,

\begin{abstract}

Magnetic kagome metals, in which topologically non-trivial band structures and electronic correlation are intertwined, have recently emerged as an exciting platform to explore exotic correlated topological phases, that are usually not found in weakly interacting materials described within the semi-classical picture of electrons. Here, via a comprehensive single-crystal neutron diffraction and first-principles density functional theory study of the archetypical topological kagome metal Mn$_3$Sn, which is also a magnetic Weyl fermion material and a promising chiral magnet for antiferromagnetic spintronics, we report the realisation of an emergent spin-density wave (SDW) order, a hallmark correlated many-body phenomenon, that is engineered by the Fermi surface nesting of topological flat bands. We further reveal that the phase transition, from the well-known high-temperature coplanar and non-collinear \textbf{k} = 0 inverse triangular antiferromagnetic order to a double-\textit{k} non-coplanar modulated incommensurate magnetic structure below $T_1$ = 280 K, is primarily driven by the SDW instability. The double-$k$ nature of this complex low-temperature magnetic order, which can be regarded as an intriguing superposition of a longitudinal SDW with a modulation wavevector \textbf{k}$_L$ and a transverse incommensurate helical magnetic order with a modulation wavevector \textbf{k}$_T$, is unambiguously confirmed by our observation of the inter-modulation high-order harmonics of the type of 2\textbf{k}$_L$+\textbf{k}$_T$. This discovery not only solves a long-standing puzzle concerning the nature of the phase transition at $T_1$, but also provides an extraordinary example on the intrinsic engineering of correlated many-body phenomena in topological matter. Due to its proximity to the room-temperature \textbf{k} = 0 chiral antiferromagnetic order, the identified multi-\textit{k} magnetic state can be further exploited for the engineering of the new modes of magnetization and chirality switching for potential applications in topological and antiferromagnetic spintronics.
  
\end{abstract}

\maketitle

\section{Introduction}

Non-trivial topology of single-electron band structures has been established as a new paradigm for quantum materials possessing large spin-orbit coupling (SOC) like that found e.g. in topological insulators and Dirac semimetals \cite{Hasan2021,Tokura2019}. These topological matters, characterized by topological invariants, for instance, Chern number in $k$-space or the skyrmion winding number in real space, until recently, are found largely in weakly interacting materials in which electronic correlation (i.e., the Coulomb repulsive interaction among the conduction electrons) plays only a minor role. On the other side, strongly correlated electron materials have long been a major source in condensed matter for the realisation of novel quantum phenomena and exotic states of matter \cite{Keimer2017, Witczak-Krempa2014}, such as high-temperature superconductivity, colossal magnetoresistance, charge and orbital ordering, and quantum spin liquid. It is thus fascinating to look into the recently emerging correlated topological materials \cite{Bernevig2022,Tokura2022}, in which both band-structure topology and strong electron correlation are intertwined, for even more exotic electronic and magnetic phenomena and novel functionalities.\\

\begin{figure*}[htbp]
	\centering
	\includegraphics[width=14cm]{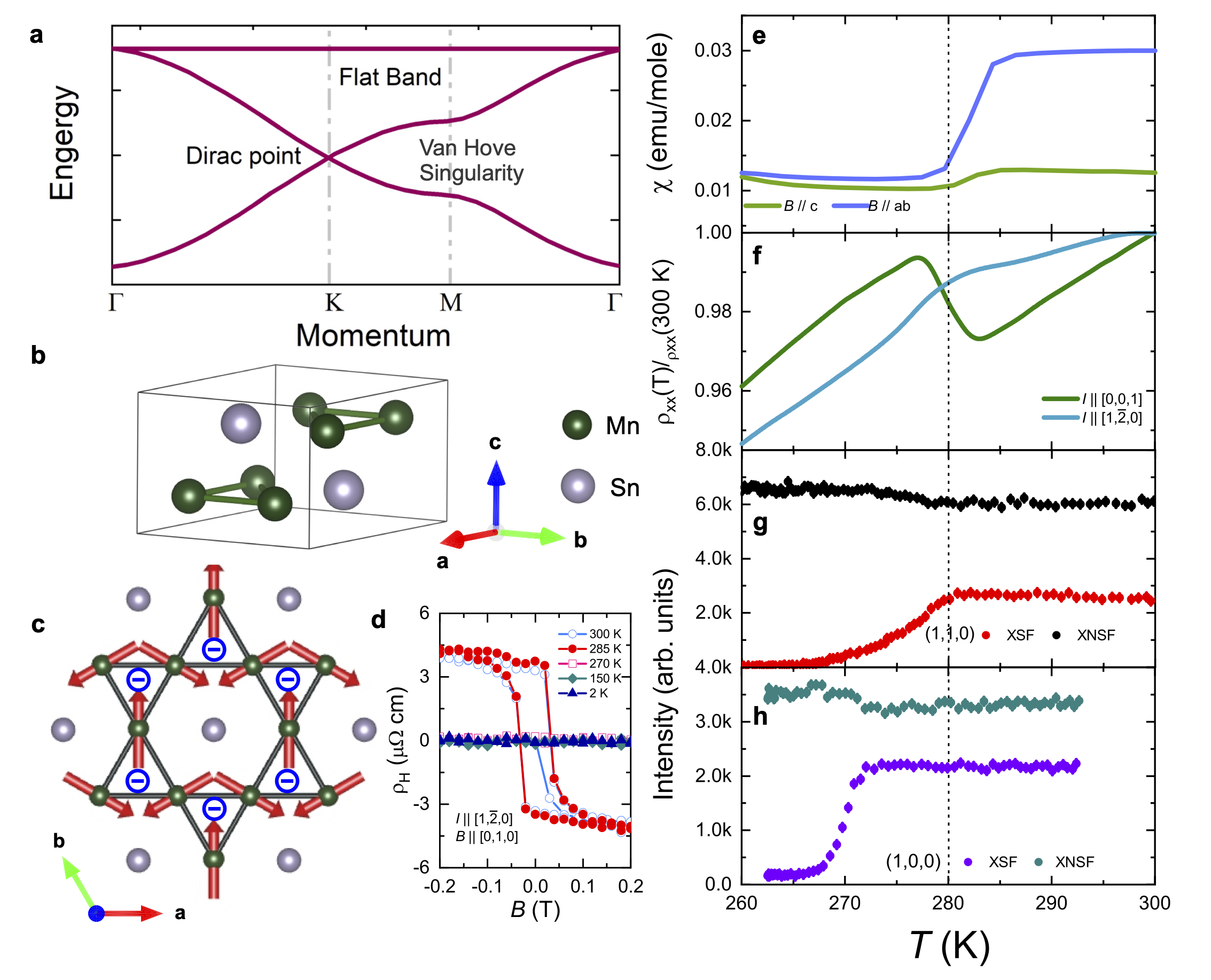}
	\caption{\label{fig:transition}
	    Band-structure topology, magnetic order and phase transitions in the kagome metal Mn$_{3}$Sn.
	    \textbf{a}, Schematics of the Dirac point at the $K$ point, the van Hove singularity at the $M$ point, and the flat band in the electronic band structures of a kagome metal.
		\textbf{b}, Crystal structure of Mn$_{3}$Sn.
		\textbf{c}, One of the possible inverse triangular antiferromagnetic structures at room temperature for Mn$_{3}$Sn, where the sign of vector spin chirality in every Mn triangle is negative.
		\textbf{d}, Field dependence of the Hall resistivity in Mn$_{3}$Sn at different temperatures. The large AHE is completely suppressed in our studied samples below the magnetic phase transition at $T_1$ = 280 K.
		\textbf{e-h}, Experimental signatures of the phase transition at $T_1$:
		(\textbf{e}), Magnetic susceptibility of Mn$_{3}$Sn measured along the $ab$ basal plane and $c$ direction with an external magnetic field \textit{H} = 0.1 T.
		(\textbf{f}), Temperature dependence of the longitudinal electric resistivity of Mn$_{3}$Sn measured with electric current along the crystallographic [0,0,1] and [1,$\overline{2}$,0] directions.
		(\textbf{g}) and (\textbf{h}), Temperature dependence of the (1,0,0) and (1,1,0) magnetic reflections measured from the $x$-spin flip ($xsf$) and $x$-non-spin flip ($xnsf$) channel, respectively, where a complete phase transition is found in our samples. The (1,0,0) and (1,1,0) reflections are measured on two single-crystal samples grown from different batches, where a difference in transition temperature could be noticed.
	}
\end{figure*}

Topological kagome metals \cite{Yin2022}, consisting of a network of corner-sharing triangles of transition-metal ions, have emerged recently as an ideal platform to explore a variety of exotic topological phases, ranging from the magnetic Weyl fermions in Co$_3$Sn$_2$S$_2$ \cite{Liu2019}, the massive Dirac fermions in the ferromagnetic Fe$_3$Sn$_2$ \cite{Ye2018}, the quantum-limit magnetic Chern phase in TbMn$_6$Sn$_6$ \cite{Yin2020b}, to the kagome superconductors \textit{A}V$_3$Sb$_5$ (\textit{A} = K, Rb and Cs) in which unconventional superconductivity coexists with chiral charge-density wave (CDW) \cite{Ortiz2020}, and the emergent CDW order in the antiferromagnetic FeGe \cite{Teng2022,Teng2023}. Due to its peculiar geometry of a kagome lattice, both the linearly dispersing and topologically protected band-crossing Dirac points, the van Hove singularities and the dispersionless flat bands could in principle be present in the electronic band structures of a kagome metal, as shown schematically in Fig. \ref{fig:transition}(a). The van Hove singularity driven electronic instability \cite{Kang2022,Teng2022,Teng2023} has indeed been suggested to be responsible for the emergence of CDW in both the kagome superconductors \textit{A}V$_3$Sb$_5$ and the kagome magnet FeGe. Meanwhile, the presence of topological flat bands would imply that the kinetic energy of electrons can be vanishingly small, and their effective mass can be exceedingly large, such that the relevant electronic states would essentially be classified as spatially localised or \textit{strongly correlated}. Therefore, flat bands can be a natural fertile ground for electronic correlation effects. Despite the recent advances in the experimental identification of flat bands in a number of magnetic kagome metals \cite{Yin2019,Kang2020a,Liu2020a,Kang2020,Lin2018,Li2021}, the quest for the flat band-driven correlated many-body phenomena remains a significant challenge. This is largely because the observed flat bands in these kagome metals are often located far away from the Fermi level $E_{F}$, while correlation effects would become prominent only if they are close to $E_{F}$.\\

Binary intermetallic compounds Mn$_{3}$\textit{A} (\textit{A} = Sn, Ge) crystallize in a hexagonal structure with space group \textit{P}6$_{3}$/\textit{mmc}, in which Mn atoms form a slightly distorted breathing-type kagome lattice in each of the \textit{z} = 1/4 and \textit{z} = 3/4 layer as shown in Fig. \ref{fig:transition}(b). Below $T_{N}$ = 420 K, Mn$_{3}$Sn exhibits a 120$^\circ$ magnetic structure with a modulation wavevector \textbf{k} = 0 and a unique negative vector spin chirality \cite{Tomiyoshi1982a,Nagamiya1982,Tomiyoshi1986,Brown1990}, also referred to as an inverse triangular antiferromagnetic order, in which the ordered Mn moments are aligned in the crystallographic \textit{ab} plane, as shown in Fig.\ref{fig:transition}(c). This unusual non-collinear antiferromagnetic structure is stabilised by the presence of a significant Dzyaloshinskii–Moriya interaction (DMI) perpendicular to the kagome lattice plane \cite{Elhajal2002,Park2018,Sticht1989}. The magnetic space group of this structure is proposed to be either $\textit{Cm}'\textit{cm}'$ or $\textit{Cmc}'\textit{m}'$ \cite{Brown1990}. As an archetypical kagome metal, Mn$_3$Sn has attracted tremendous interests largely owing to the recent observations of a strikingly large anomalous Hall effect (AHE) \cite{Nakatsuji2015} and a range of the related anomalous transport properties \cite{Ikhlas2017,Higo2018,Kimata2019,Li2023} at room temperature, as well as the realisation of exotic Weyl fermions \cite{Kubler2017,Yang2017,Kuroda2017} in this material. While as suggested in various theories \cite{Chen2014,Kubler2014,Suzuki2017,Busch2020} these anomalous transport properties are intimately linked to Berry curvature in $k$-space that is induced by this coplanar and non-collinear antiferromagnetic order in a kagome lattice, they are considered also as an experimental fingerprint of the topologically protected Weyl nodes residing near $E_{F}$. The interplay between non-collinear antiferromagnetism, Berry curvature of the electronic structures and Weyl fermions in Mn$_3$Sn is believed to play a crucial role in these anomalous transport properties \cite{Chen2021} as well as in their remarkable room-temperature tunability, for instance, by electric currents \cite{Tsai2020,Higo2022}, spin-orbit torque \cite{Takeuchi2021}, and uniaxial strain \cite{Ikhlas2022}. In this regard, Mn$_3$Sn is also hailed as a very promising material for topological and antiferromagnetic spintronics applications \cite{Smejkal2018}. Furthermore, the recent observations of the many-body Fano resonance \cite{Zhang2020} and the Kondo effect \cite{Khadka2020} have indeed suggested that Mn$_{3}$Sn is a strong candidate material to explore topology-driven correlated many-body phenomena.\\ 

The hexagonal structure of the bulk material Mn$_{3}$Sn is stable only with a small amount of excess Mn that would intrinsically substitute at the Sn sites \cite{Tomiyoshi1982,Ikhlas2020}, therefore, depending on the exact chemical composition, the physical properties of Mn$_{3}$Sn may exhibit small sample dependence among samples prepared via different crystal growth methods. Since the electronic states at $E_{F}$ are occupied solely by the Mn 3$d$ electrons, a small amount of Mn doping would cause a sizeable shift of the chemical potential relative to the $E_{F}$ comparing to the stoichiometric case. For instance, a 1\% Mn-doping in Mn$_3$Sn may shift up the chemical potential by about 6 meV from $E_{F}$ \cite{Kuroda2017,Chen2021}. Therefore, this intrinsic albeit small doping effect due to excess Mn can have a substantial impact on the Fermi level of the electronic band structures, if acted in a controllable way, it may even be used for an intrinsic tuning of $E_{F}$ in this material. While the antiferromagnetic phase transition at $T_N$ appears a very robust feature across different sources of samples, a second phase transition at around $T_1$ = 280 K, that occurs within the inverse triangular antiferromagnetic ordered parent phase, to a possible low-temperature spiral magnetic phase, has also been hinted in the scattered literatures over the past decades \cite{Cable1993,Cable1994,Sung2018,Park2018,Ikhlas2020}. Intriguingly, an accompanied complete suppression of the large AHE below $T_1$ was also observed in the samples grown by the self-flux method recently \cite{Sung2018,Yan2019}, suggesting a significant and simultaneous change in both magnetic structure and transport properties. This also offers a new dimension for the switching of these remarkable anomalous transport properties in Mn$_{3}$Sn via a small change in temperature. A recent study on sample dependence further suggests that the phase transition at $T_1$ would emerge only when the chemical composition of a sample is close to the ideal stoichiometry \cite{Ikhlas2020}. However, the underlying mechanism for the occurrence of this highly unusual phase transition as well as the exact magnetic structure below $T_1$ have yet to be established. \\

In this article, we report a comprehensive study of the complex magnetic order and electronic band structures of the archetypical topological kagome metal Mn$_3$Sn via both single-crystal neutron diffraction experiments and first-principles density functional theory (DFT) calculations. A key finding of this study is the observation of an emergent spin-density wave (SDW) order and the accompanied exotic multi-\textit{k} magnetic state at low temperatures in this compound. Based on neutron polarisation analysis and magnetic structure refinement, we found that Mn$_3$Sn transforms, from a high-temperature coplanar and non-collinear \textbf{k} = 0 inverse triangular antiferromagnetic order, to a double-\textit{k} non-coplanar modulated incommensurate magnetic structure below $T_1$, which can be regarded as an intriguing superposition of a longitudinal SDW with a modulation wavevector \textbf{k}$_L$, and a transverse incommensurate helical magnetic order with a modulation wavevector \textbf{k}$_T$. The nature of this complex low-temperature magnetic order is unambiguously demonstrated by the observation of the inter-modulation high-order harmonics of the type of 2\textbf{k}$_L$+\textbf{k}$_T$. Furthermore, a flat band along the high-symmetry \textit{K}-\textit{M}-\textit{K} direction near $E_{F}$ is revealed by our DFT band-structure calculations. We found that the parallel sections of this flat band on the Fermi surface form a perfect nesting condition that matches to the modulation wavevector of the observed incommensurate magnetic structures. We thus argue that the magnetic phase transition at $T_1$ is primarily driven by a SDW instability that is associated to the Fermi-surface nesting of flat bands, also a hallmark correlated electron phenomenon. Our DFT calculations also quantitatively demonstrate that the small intrinsic doping effect due to excess Mn is indeed responsible for shifting $E_{F}$ near to this flat band, and thus paves the way for the emergence of rich Fermi surface-mediated many-body correlation effects. The discovery of a topological flat band-engineered SDW in Mn$_{3}$Sn not only solves a long-standing puzzle concerning the nature of the phase transition around $T_1$, but also provides an extraordinary and also a very rare example on the intrinsic engineering of correlated many-body phenomena in topological kagome metals. Given that the formation of SDW would considerably alter the electronic structure near $E_{F}$, for instance, by opening a small gap, therefore, potential impact on the Weyl fermions from SDW as well as a possible interplay between them would become a fascinating topic to explore in the future. Furthermore, the identified multi-\textit{k} magnetic states in such a prototypical example of magnetic topological materials could offer new possibilities for the engineering of the new modes of magnetization and chirality switching in Mn$_{3}$Sn and the related kagome chiral antiferromagnets for potential applications in topological and antiferromagnetic spintronics. \\

\section{Materials and Methods}

High-quality Mn$_{3}$Sn single crystals were grown from the molten self-flux methods \cite{Duan2015,Sung2018,Yan2019}. The starting materials consisting of Mn pieces (99.95\% purity, Alfa Aesar) and Sn granules (99.99\% purity, Chempur) were loaded in an Al$_2$O$_3$ crucible with some quartz wools on its top as a filter, and the crucible was then placed in a quartz ampoule that was sealed in vacuum afterwards. The quartz ampoule was heated to 1100 \textcelsius\ in 10 hours in a tube furnace, subsequently dwelt at 1100 \textcelsius\ for 20 hours, and then cooled down to 900 \textcelsius\ at a rate of 1 \textcelsius/hour. When the temperature decreased to 900 \textcelsius\ the ampoule was quickly transferred to a centrifuge to separate crystals from the Sn flux. In some batches, some crystals are not fully separable from the flux, and the excess Sn flux could then be removed mechanically. The chemical composition of the as-grown single crystals was checked via energy dispersive X-ray analysis (EDX). The crystalline quality and orientation of the selected crystals were carefully examined by X-ray Laue. The magnetic properties of a number of Mn$_{3}$Sn crystals grown from different batches were measured between 1.8 K and 300 K on a SQUID magnetometer system (from Quantum Design). Both the longitudinal electric resistivity and Hall effect measurements were carried out from 2 K and 300 K on a PPMS system (from Quantum Design).\\

A range of comprehensive single-crystal neutron diffraction experiments were carried out on our Mn$_{3}$Sn samples. The non-polarised single-crystal neutron diffraction experiment was performed at the hot-neutron 4-circle diffractometer HEiDi \cite{Meven2015} at the Heinz Maier-Leibnitz Zentrum (MLZ), with an incident wavelength of 0.87 \AA. For the precise determination of nuclear structure and chemical compositions of the studied sample, in total 707 nuclear reflections were measured in the incommensurate magnetic phase at 210 K, in which nuclear and magnetic reflections are well separated. Furthermore, 612 reflections, including 142 nuclear and 470 magnetic ones, were measured for the refinement of the magnetic structure in the lock-in phase at 225 K. The determination of the coplanar and non-collinear \textbf{k} = 0 antiferromagnetic structure was based on the 1343 reflections collected in the inverse triangle antiferromagnetic phase at 300 K. A further non-polarised single-crystal neutron diffraction experiment was undertaken for the detailed measurements of both temperature and magnetic field dependence at the lifting-counter thermal-neutron diffractometer D23 with a 6 T vertical-field magnet and an incident wavelength $\lambda_i$ = 2.36 \AA. For this experiment, the ($h$,0,$l$) reciprocal plane of the studied crystal is aligned in the horizontal scattering plane, so that the applied magnetic field is in parallel to the [$\overline{1}$,2,0] direction of this hexagonal lattice system. The polarized neutron diffraction experiments were carried out at the cold-neutron polarised spectrometer DNS \cite{Su2015} (with $\lambda_i$ = 4.74 \AA) at MLZ, and at the cold-neutron triple-axis spectrometer IN12 at the Institut Laue–Langevin (ILL) (with k$_i$ = k$_f$ = 2 \AA$^{-1}$). A neutron velocity selector is employed for the filtering out of higher-harmonics at both DNS and IN12. A Helmholtz XYZ-coil system and a zero-field CRYOPAD system were used for the $xyz$ polarisation analysis at DNS and IN12, respectively. The flipping ratio of the typical nuclear reflections obtained at both instruments on the studied Mn$_{3}$Sn samples is in the range of 20-25. For the experiments at DNS, the thoroughness of the magnetic phase transition at $T_1$ was carefully checked in a number of single crystals grown from different batches, including both ($h$,0,$l$) and ($h$,$h$,$l$) oriented in the horizontal scattering plane. The crystals that show a complete transition at $T_1$ were chosen for the polarisation analysis study. For the IN12 experiment, only the crystal oriented in ($h$,0,$l$) was measured. No noticeable neutron depolarisation could be seen in the studied temperature range of the neutron polarisation analysis measurements. \\ 

The electronic structure of Mn$_3$Sn was calculated based on the first-principles density functional theory with the generalized-gradient approximation (GGA) in the form of Perdew-Burke-Ernzerhof \cite{Perdew1996}. The accurate frozen-core full-potential projector augmented wave method \cite{Blochl1994}, as implemented in the Vienna ab initio simulation package (VASP) \cite{Kresse1993,Kresse1996}, was used. The fully relativistic projector augmented potentials are adopted in order to include the spin-orbit coupling. The plane-wave energy cut-off of 300 eV and a Monkhorst-Pack $k$-point mesh of 9$\times$9$\times$9 were used for the self-consistent field calculations. After having obtained the converged charge density, the maximally localized Wannier functions were constructed by projecting onto the $s$, $p$, and $d$ orbitals of Mn atom as well as onto $s$ and $p$ orbitals of Sn atom, using the WANNIER90 package \cite{Pizzi2020}. The Fermi surface is then calculated on a denser $k$-point mesh of 51$\times$51$\times$51 by the means of Wannier interpolation.\\

\section{Results}

\subsection{Phase transition at $T_1$} 

The phase transition at $T_1$ in our flux-method grown single crystals has been carefully examined via the measurements of magnetic and transport properties and polarised neutron diffraction. As shown in Fig.\ref{fig:transition}(d), the large AHE in the magnetic parent phase is completely vanished below $T_1$ in our samples and does not recover at lower temperatures. The magnetic susceptibility measured along both the $c$ axis and crystallographic \textit{ab} plane undergoes a drastic drop at around $T_1$ (see Fig.\ref{fig:transition}(e)). This drastic change can also be confirmed in the normalised longitudinal resistivity $\rho_{xx}(T)/\rho_{xx}$(300 K) measured with electric current applied along the [0,0,1] and [$\overline{1}$,2,0] directions, respectively, as shown in Fig.\ref{fig:transition}(f), in which a notable jump along the [0,0,1] direction could be seen clearly at $T_1$. These observations thus indicate that the phase transition at $T_1$ is clearly associated with a substantial and also collective change in electronic, magnetic and topological properties. In addition, given that the inverse triangular antiferromagnetic order at room temperature is \textbf{k} = 0 type, the nuclear reflections coexist with the magnetic ones at the same $q$ position. With the help of polarised neutron scattering, the magnetic scattering cross-section can be separated from the contribution of the nuclear coherent scattering cross-section, i.e., the $x$-spin flip channel ($xsf$) uniquely for magnetic reflections, and the $x$-non-spin flip channel ($xnsf$) only for nuclear reflections. As shown in Fig.\ref{fig:transition}(g) and (h), the temperature dependence of the polarised neutron diffraction measurements shows the complete vanishing of magnetic intensities of the (1,0,0) and (1,1,0) reflections in the $xsf$ channel, demonstrating the transition at $T_1$ in the studied samples is indeed thorough and complete. Further discussions about this issue can be found in the Appendix. \\

\begin{figure}[ttbp]
	\centering
	\includegraphics[width=8cm]{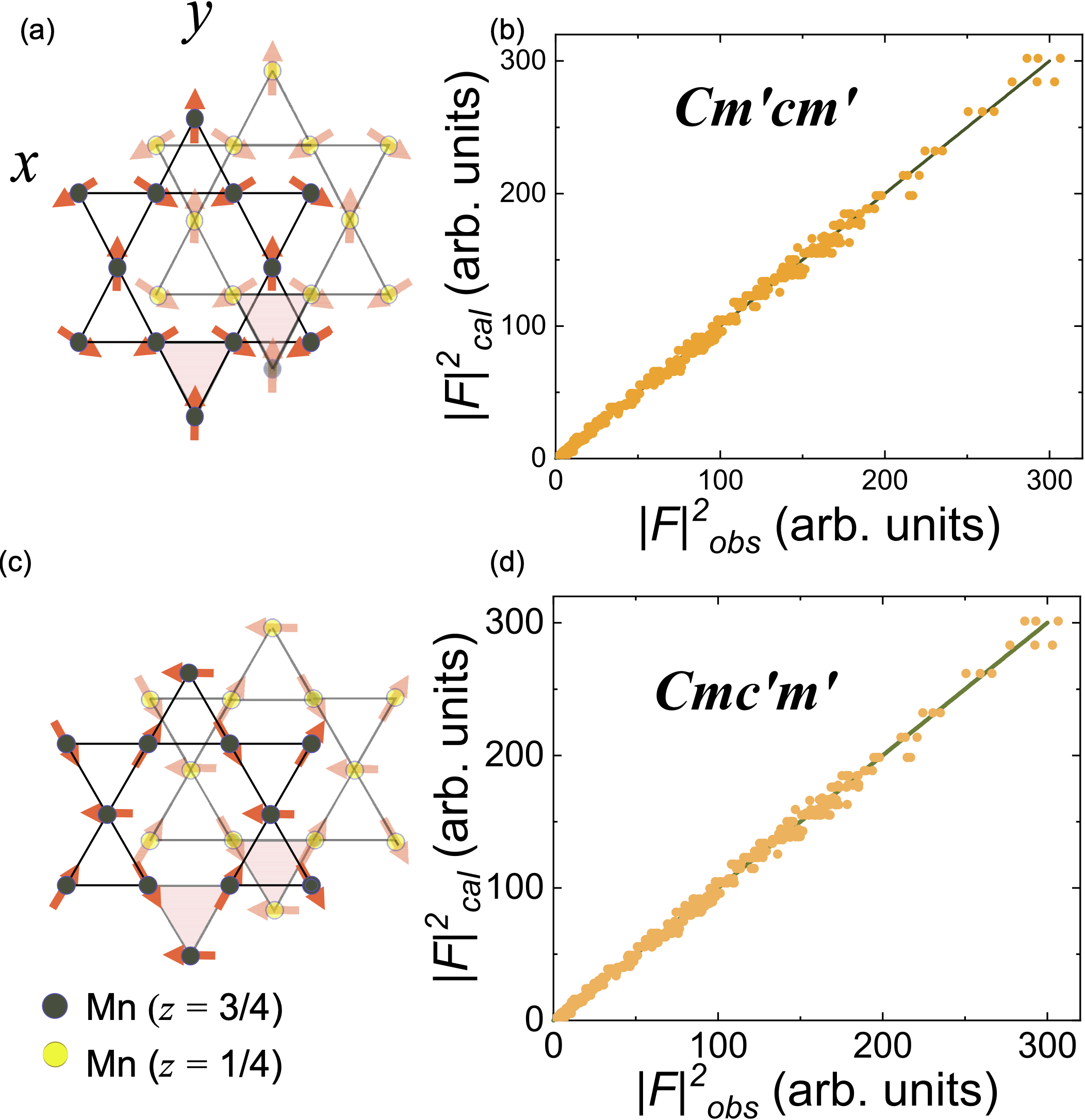}
	\caption{\label{fig:mag_str_300K}
		Two possible magnetic structure models and the respective calculated squared structure factors versus the measured ones based on the Jana2006 refinement. \textbf{a-b}, \textit{Cm'cm'}. \textbf{c-d}, \textit{Cmc'm'}.
	}
\end{figure}

\subsection{Magnetic structure at 300 K}

To confirm the magnetic structure of Mn$_{3}$Sn above $T_1$, we carried out a detailed single-crystal neutron diffraction experiments at 300 K at HEiDi. In total, 1343 reflections were collected for the structural refinement with Jana2006 \cite{Petricek2014}. It is known that the room-temperature magnetic structure of Mn$_{3}$Sn is a type of $\mathbf{k}$ = (0,0,0) antiferromagnetic order, in which the magnetic reflections appear at the same position as the nuclear ones. We used the atomic occupancies that are already obtained  at 210 K (see Tab.\ref{tab:struture} in the Appendix A) for the refinements. Based on the previous studies \cite{Brown1990}, there are two possible magnetic space groups proposed for the room-temperature magnetic structure, namely, $Cmc'm'$ or $Cm'cm'$. The corresponding magnetic structure models and the respective refinement results are shown in Fig.\ref{fig:mag_str_300K}(a-d) and Tab.\ref{tab:mag_300K}. The calculated structure factors are quite consistent to the measured values for both models. The ordered magnetic moment at 300 K is determined to be 2.54(8) $\mu_B$ and 2.53(10) $\mu_B$ respectively for each model, which are close to 3.00(1) $\mu_B$ obtained in an earlier study \cite{Brown1990}. To distinguish between these two possible models with non-polarised single-crystal neutron diffraction is challenging because of the presence of multiple magnetic domains at zero field. This challenge could not even be completely overcome in an earlier study via spherical neutron polarisation analysis on a magnetized sample performed by P.J. Brown \textit{et al.} \cite{Brown1990}. Note that in a recent spherical neutron polarisation analysis study on a closely related compound Mn$_{3}$Ge, it has been shown that the magnetic space group \textit{Cm'cm'} is favoured for the description of the ground state magnetic structure \cite{Soh2020}.

\begin{table}[htbp]
	\caption[Mn$_3$Sn magnetic refinement results]{\label{tab:mag_300K}
		The determined magnetic structures of Mn$_{3}$Sn at 300 K with a magnetic propagation vector $\mathbf{k}$ = (0,0,0).
	}
	\begin{center}
%		\begin{tabular}{lp{2cm}p{2cm}p{2cm}p{2cm}p{2cm}p{2cm}p{2cm}l}
		\begin{tabular}{cp{0.8cm}p{0.6cm}p{1.5cm}p{1.3cm}p{0.8cm}p{0.8cm}p{1cm}c}
			\hline
			\hline
			\multicolumn{9}{c}{\textrm{Magnetic space group model 1: $Cm'cm'$}}\\
			\hline
			&Atom&Site&$x$&$y$&$z$&$Occ.$&$U_{iso}$ (\AA$^2$)& \\
			\hline
			&Mn1&$6h$&0.8387(1)&0.6775&0.25&1.016&0.00615(13)& \\
			&Mn2&$6h$&-0.6775&0.1612(1)&0.25&1.016&0.00615(13)& \\
			&Sn&$2c$&0.3333&0.6667&0.25&0.333&0.00462(12)& \\
			\hline
			&\multicolumn{8}{l}{\textrm{$R(obs)$ = 2.27\%,  $R_w(obs)$ = 2.72\%,  $R_{w}(all)$ = 3.05\%}}\\
			\hline
			&\multicolumn{8}{l}{\textrm{$M_{Mn}$ = 2.54(8) $\rm \mu_B$}}\\
			\hline
			\hline
			\multicolumn{9}{c}{\textrm{Magnetic space group model 2: $Cmc'm'$}}\\
			\hline
			&Atom&Site&$x$&$y$&$z$&$Occ.$&$U_{iso}$ (\AA$^2$)& \\
			\hline
			&Mn1&$6h$&0.8386(1)&0.6773&0.25&1.016&0.00617(13)& \\
			&Mn2&$6h$&-0.6773&0.1614(1)&0.25&1.016&0.00617(13)& \\
			&Sn&$2c$&0.3333&0.6667&0.25&0.333&0.00467(13)& \\
			\hline
			&\multicolumn{8}{l}{\textrm{$R(obs)$ = 2.30\%,  $R_w(obs)$ = 2.75\%,  $R_w(all)$ = 3.08\%}}\\
			\hline
			&\multicolumn{8}{l}{\textrm{$M_{Mn}$ = 2.53(10) $\rm \mu_B$}}\\
			\hline
			&\multicolumn{8}{l}{\textrm{$a$ = $b$ = 5.67 \AA,  $c$ = 4.53 \AA}}\\
			\hline
			&\multicolumn{8}{l}{$Occ.$: atomic occupation.}\\
			&\multicolumn{8}{l}{$U_{iso}$: isotropic atomic displacement parameters.}\\
			\hline
			\hline
		\end{tabular}
	\end{center}
\end{table}

\begin{figure*}[htbp]
	\centering
	\includegraphics[width=14cm]{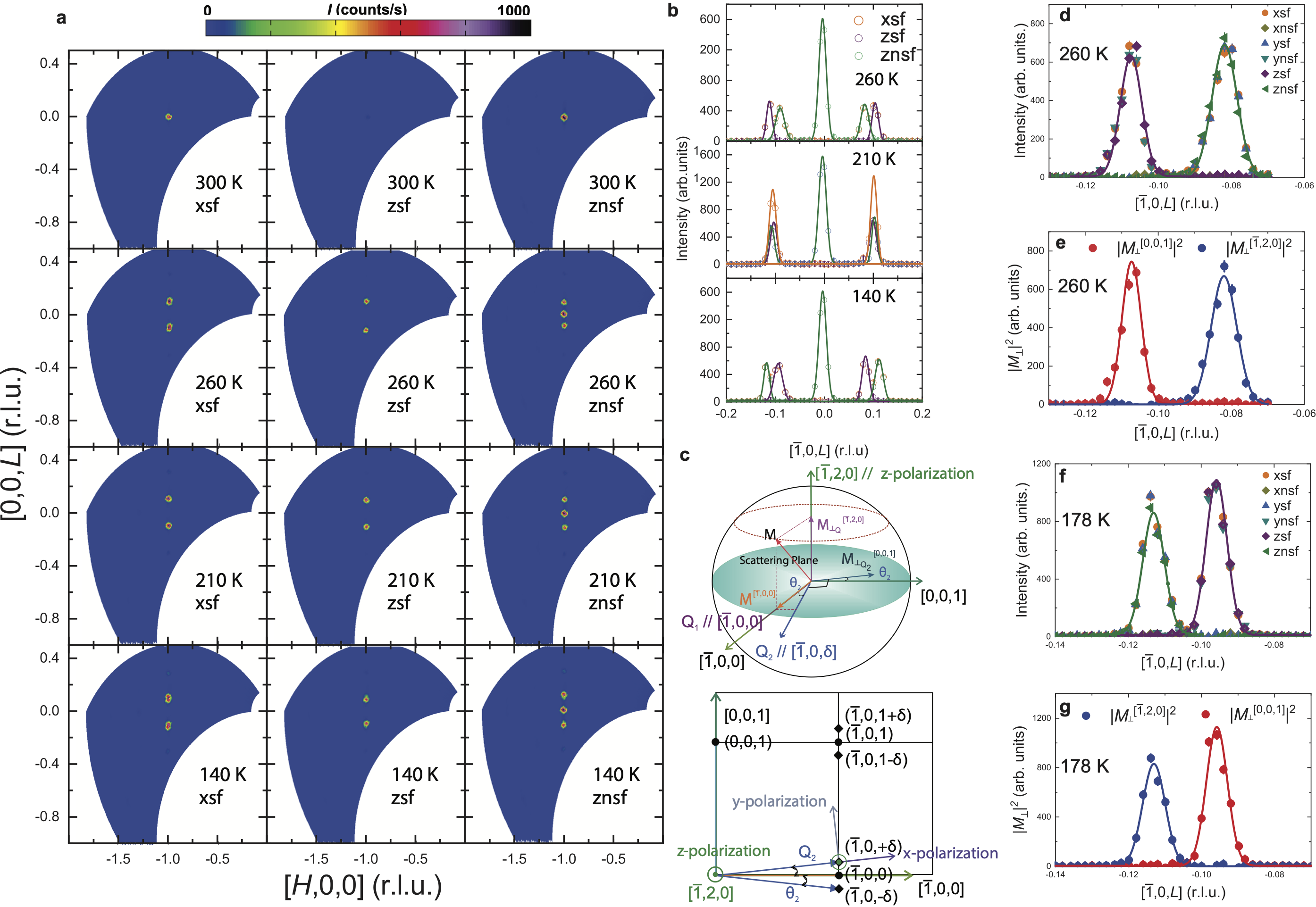}
	\caption{\label{fig:polarised}
	    Complex double-$k$ incommensurate magnetic orders in Mn$_{3}$Sn unravelled by polarised neutron diffraction.
		\textbf{a}, The reciprocal-space mappings of the $xsf$, $zsf$ and $znsf$ scattering channels measured at DNS on an as-grown single crystal of Mn$_{3}$Sn at 300, 260, 210 and 140 K, respectively.
		\textbf{b}, The selected line-cuts along the [$\overline{1}$,0,\textit{L}] direction are obtained from the respective intensity maps in (\textbf{a}). It is clear that the two adjacent satellite reflections in the $xsf$ channel could be well separated respectively in the $zsf$ and $znsf$ channels, even at 210 K, where they have a common modulated wavevector. 
		\textbf{c}, The scattering geometry for polarisation analysis at DNS. The horizontal scattering plane is the ($h$,0,$l$) reciprocal plane, where the magnetic interaction vector $\mathbf{M}$ in the ($h$,$k $,0) plane can be decomposed into $\mathbf{M}^{[\overline{1},2,0]}$ and $\mathbf{M}^{[\overline{1},0,0]}$. For the reflection ($\overline{1}$,0,0), the scattering vector Q$_{1}$ is parallel to the [$\overline{1}$,0,0] direction. In this case, only the magnetic component along the [$\overline{1}$,2,0] or [0,0,1] direction can be probed by polarised neutron scattering, which will contribute to the intensities in the $znsf$ and $zsf$ channels, respectively. For the magnetic satellite reflections, assuming Q$_{2}$ is the scattering vector and $\theta_{2}$ is the angle between the scattering vector Q$_{2}$ and the ($h$,$k$,0) plane. In the $zsf$ channel, the magnetic intensity can thus be expressed as $|\mathbf{M}^{[\overline{1},0,0]}|^{2}\sin^{2}\theta + |\mathbf{M}^{[0,0,1]}|^{2}\cos^{2}\theta$. If the background and incoherent scattering are negligible, in the $znsf$ channel, the intensity is still from the magnetic component in ($h$,$k$,0) plane, which is $|\mathbf{M}^{[\overline{1},2,0]}|^{2}$.
		\textbf{d-f}, The representative polarisation dependent $Q$-scans along the [$\overline{1}$,0,\textit{L}] direction measured at IN12 with higher resolution in the $xsf$, $xnsf$, $ysf$, $ynsf$, $zsf$ and $znsf$ channels at 260 K (\textbf{d}) and 178 K (\textbf{f}), respectively. The derived squared components of magnetic interaction vector $|M_{\perp}^{[0,0,1]}|^{2}$ and $|M_{\perp}^{[\overline{1},2,0]}|^{2}$ at 260 K and 178 K are shown in (\textbf{e}) and (\textbf{g}), respectively. 
	}
\end{figure*}

\subsection{Unravelling complex magnetic order below $T_1$ via polarised neutron diffraction}
   
To reveal the magnetic structures below $T_{1}$, a series of polarised and non-polarised neutron diffraction experiments were carried out on our Mn$_{3}$Sn single crystals. Neutron polarisation-resolved reciprocal space intensity maps were measured at various temperatures in the ($h$,0,$l$) scattering plane as shown in Fig.\ref{fig:polarised}(a). For longitudinal neutron polarisation analysis employed in our measurements, we define three orthogonal directions in a Cartesian coordination system: $x$ is parallel to the scattering vector $\mathbf{Q}$, $z$ is perpendicular to the scattering plane, and $y$ is determined by the right-hand rule and also in the scattering plane. At 300 K i.e. in the inverse triangle antiferromagnetic parent phase, a sharp magnetic peak ($\overline{1}$,0,0) is observed in the $xsf$ and $znsf$ channels, but is absent in the $zsf$ channel. When the sample is cooled down to 260 K i.e. below $T_1$, ($\overline{1}$,0,0) disappears in the $xsf$ channel, and a set of new satellite peaks emerge at around (1,0,$\pm$0.1) in the $xsf$, $zsf$ and $znsf$ channels. The one-dimensional profile cuts from the reciprocal space intensity maps at 260, 210 and 140 K are plotted in Fig.\ref{fig:polarised}(b), in which two types of incommensurate magnetic satellite peaks are evident in the $xsf$ channel below $T_{1}$, and surprisingly, they are well separable between the $zsf$ and $znsf$ channels, even at 210 K where there is only a single satellite peak at ($\overline{1}$,0,$\pm0.1$). This observation is verified by further high-resolution polarised neutron diffraction measurements at 260 and 178 K as shown respectively in Fig.\ref{fig:polarised} (d) and (f). Among six measured polarisation cross-section channels at each temperature, these two different types of magnetic satellite peaks that both appear in the $xsf$ channel are indeed well separated among the $ynsf$, $zsf$ and $ysf$, $znsf$ channels, respectively. This finding suggests that neutron polarisation analysis can provide a unique way for \textit{labelling} multiple magnetic satellite peaks, thus paving the way for a systematic study of the complex magnetic structures below $T_{1}$ in Mn$_{3}$Sn.\\

Based on the experimental scattering geometry (see Fig.\ref{fig:polarised}(c)), if both the background and incoherent scattering contributions are negligible, the polarised neutron scattering cross-sections can be expressed in the following equations with respect to the magnetic interaction vectors $\mathbf{M}_{\perp}$,

\begin{equation}
\begin{aligned}
&\sigma_{z}^{sf}=\sigma_{x}^{sf}-\sigma_{y}^{sf}=\mathbf{M}_{\perp y}^{2}\\
& =|\mathbf{M}^{[\overline{1},0,0]}|^{2}\sin^{2}\theta  + |\mathbf{M}^{[0,0,1]}|^{2}\cos^{2}\theta
\end{aligned}
\end{equation}

\begin{equation}
\begin{aligned}
&\sigma_{y}^{sf}=\sigma_{z}^{nsf}-\sigma_{x}^{nsf}=\mathbf{M}_{\perp z}^{2}=|\mathbf{M}^{[\overline{1},2,0]}|^{2}
\end{aligned}
\end{equation}

\noindent where $\sigma_{i}^{j}$ ($i = x,y,z$, $j = sf, nsf$) represents the spin flip ($sf$) or non-spin flip ($nsf$) cross-section in polarisation $i$, and $\theta$ is the angle between the scattering vector $\mathbf{Q}$ and the [1,0,0] direction. $\mathbf{M}_{\perp y}$ and $\mathbf{M}_{\perp z}$ are the components of the magnetic interaction vector $\mathbf{M}_{\perp } = \langle\mathbf{M}\times \mathbf{Q} \times \mathbf{M}\rangle$ along the polarisation direction $y$ and $z$, respectively. At 300 K, for the reflection ($\overline{1}$,0,0) the scattering vector $\mathbf{Q}_{1}$ is in the ($h$,$k$,0) plane and parallel to the [$\overline{1}$,0,0] direction, so that $\theta_{1}$ equals to zero. Given that the ($\overline{1}$,0,0) reflection is observed only in the $xsf$ and $znsf$ channels, it can thus be concluded that the ordered magnetic moments lie only in the $ab$ plane, in agreement with the previous reports \cite{Brown1990,Tomiyoshi1986}. For the incommensurate magnetic satellite reflections in the modulated magnetic phase below $T_1$, the scattering vector $\mathbf{Q}_{2}$ is not in the ($h$,$k$,0) plane, and, for instance, $\theta_{2}$ is $\tan^{-1}(((2\pi/c)\times0.1)/(2\pi/(\sqrt{3}a/2)))$ = 6.22$^\circ$ for the satellite reflections (1,0,$\pm0.1$), we have then sin$^{2}\theta_{2}$ = 0.01, which means that only about 1\% of the magnetic component along the [1,0,0] direction would contribute to $\sigma_{z}^{sf}$, and that its contribution is essentially negligible in our analysis since it is within the error bar of our measurement. As a result of this analysis, as shown in Fig.\ref{fig:polarised} (e) and (g), the magnetic satellite reflections around (1,0,$\pm$0.1) could be well decomposed into two orthogonal contributions: $\mathbf{M}_{\perp y}^{2}= |\mathbf{M}^{[0,0,1]}|^{2}$ and $\mathbf{M}_{\perp_z}^{2}=|\mathbf{M}^{[\overline{1},2,0]}|^{2}$. This would indicate that the modulated magnetic phase below $T_1$ is not a spiral order as previously suggested \cite{Cable1993,Cable1994,Sung2018,Park2018}, but possesses an intriguing non-coplanar magnetic structure. From now on, we denote the magnetic modulation that is associated with the observed magnetic satellite reflections in the $znsf$ ($zsf$) channel as \textbf{k}$_T$ (\textbf{k}$_L$), since it is transverse ($longitudinal$) to the magnetic modulation wavevector. \\

\begin{figure}[htbp]
	\centering
	\includegraphics[width=8cm]{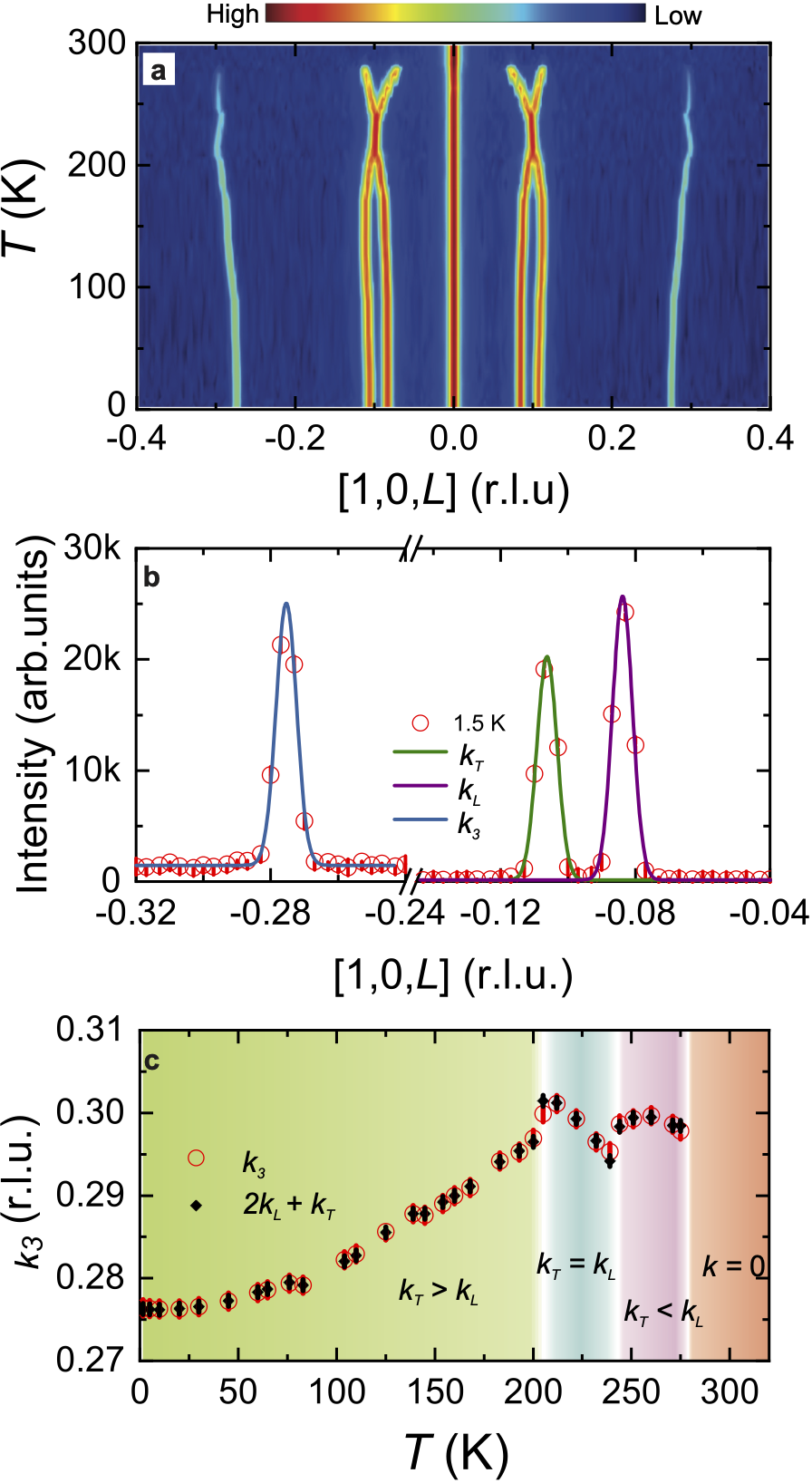}
	\caption{\label{fig:3rd}
	    \textquotedblleft Dancing in Pairs\textquotedblright : Intriguing temperature evolution of the satellite reflections in the double-$k$ incommensurate magnetic phase below $T_{1}$ in Mn$_{3}$Sn.
		\textbf{a}, Temperature dependence of the $Q$-scan along the [1,0,$L$] direction measured at D23, where two sets of magnetic satellite reflections (i.e., \textbf{k}$_T$, \textbf{k}$_L$ ) and the high-order harmonics \textbf{k}$_{3}$ emerge below $T_{1}$ = 280 K.
		\textbf{b}, These $Q$-scans from (\textbf{a}) can be well fitted with Gaussian functions. Note that the intensity of the high-order harmonic \textbf{k}$_{3}$ in the plot has been multiplied by a factor of 10, and that only the small $Q$-regions around the peak positions are shown in the plot. 
		\textbf{c}, As shown in its temperature dependence, it can be unambiguously concluded that the high-order harmonic \textbf{k}$_{3}$ can only be indexed as the inter-modulation high-order harmonics of 2\textbf{k}$_L$+\textbf{k}$_T$. Furthermore, based on this temperature dependence, the double-$k$ incommensurate magnetic ordered phase can be divided into four distinct regimes according to the propagation wavevectors: \textbf{k} = 0, \textbf{k}$_T$ \textless\ \textbf{k}$_L$, \textbf{k}$_T$ = \textbf{k}$_L$, and \textbf{k}$_T$ \textgreater\  \textbf{k}$_L$.
	}
\end{figure}

\begin{figure*}[htbp]
	\centering
	\includegraphics[width=14cm]{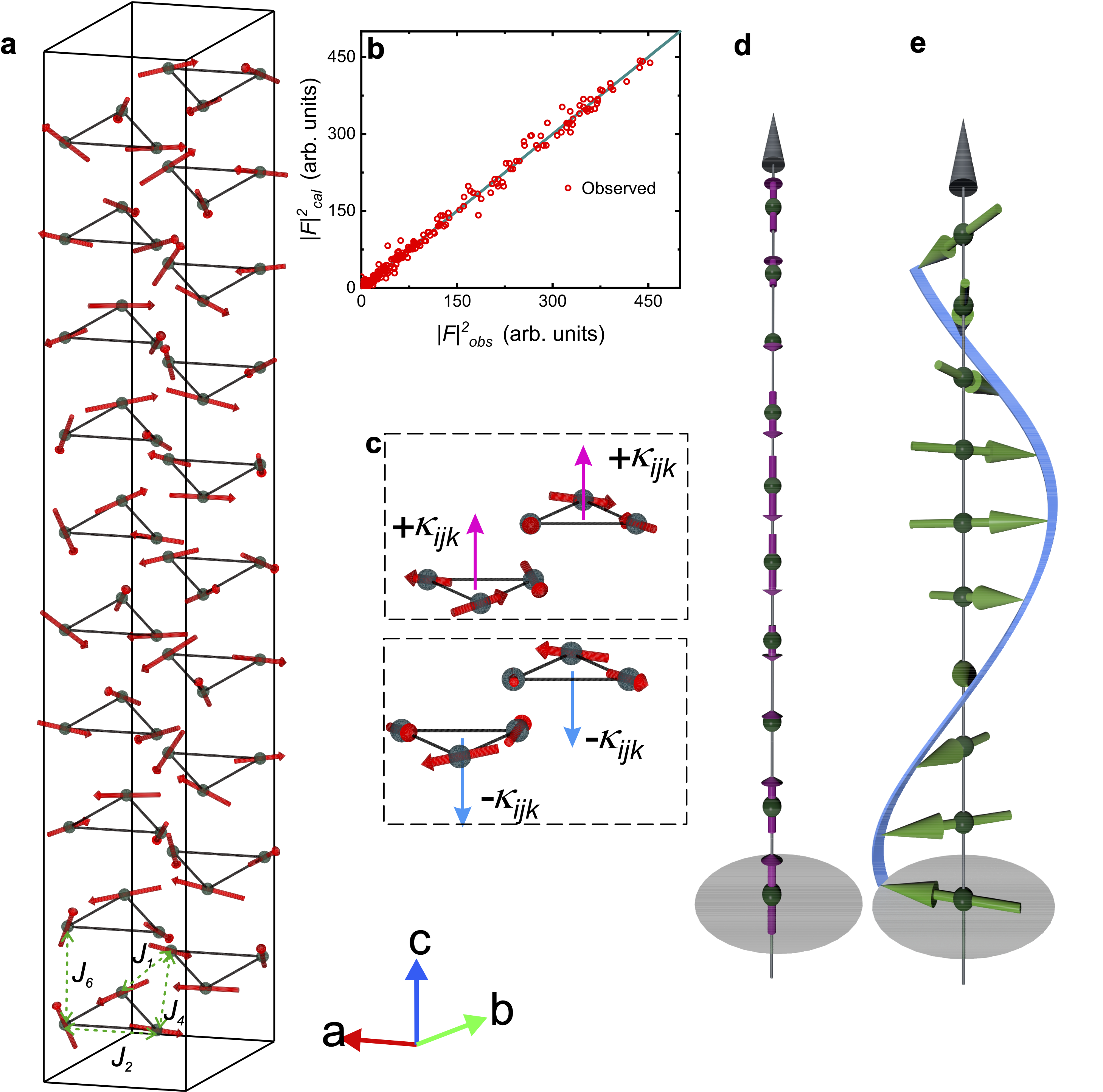}
	\caption{\label{fig:mag_str}
		The magnetic structure in the lock-in phase of Mn$_{3}$Sn that is characterised by the magnetic modulation wavevector \textit{k} = (0,0,0.1).
		\textbf{a}, The magnetic supercell, determined based on the neutron diffraction data taken at 225 K at HEiDi, has a long-periodicity modulation and a non-coplanar magnetic configuration. The magnetic configuration of each layer repeats itself after ten layers. For each layer $l_{ci}$, the magnetic moment of the layer $l_{(c+5)i}$ is reversed by time reversal symmetry. As shown in (\textbf{c}), the 3rd layer has positive scalar spin chirality while the 8th layer has negative scalar spin chirality, leading to the vanishing of the global scalar spin chirality. The AHE could not be realised in the materials with such a symmetry. The magnetic exchange interactions over different neighbour-distances are shown as the dashed green lines in the plot. The definition of $J_i$ ($i = 1, 2, 4, 6$) is based on the reference \cite{Park2018}.
		\textbf{b}, A plot of the squared magnetic structure factors after the refinement (i.e., the calculated $vs.$ observed ones).		 
		\textbf{d-e}, The magnetic structure can be seen as the superposition of: (\textbf{d}), a longitudinal spin density wave along the $c$ direction; (\textbf{e}), and the elliptical spiral magnetic structure in the $ab$ plane.
	}
\end{figure*} 

\subsection{Temperature dependence of the magnetic order below $T_1$}

Having established the way to resolve the different magnetic satellite refections via neutron polarisation analysis, we turn to the evolution of the magnetic structures as a function of temperature. As shown in the [1,0,$L$] Q-scans over a wide temperature range in Fig.\ref{fig:3rd}(a), initially, two pairs of magnetic satellite reflections centred around (1,0,0) emerge below $T_{1}$ = 280 K, then in a fascinating way, they merge or lock-in to a single one at (1,0,$\pm$0.1) from 240 to 200 K, and eventually split to two again below 200 K. Surprisingly, additional satellite peaks are evident at around (1,0,$\pm$0.3), which are almost 1-2 orders of magnitude weaker than the main satellite peaks and seemingly resemble the third-order harmonics of magnetic modulations \textbf{k}$_{3}$. By fitting the observed satellite peaks at each temperature with a Gaussian function (as shown in Fig.\ref{fig:3rd}(b)), the modulation wavevectors of \textbf{k}$_L$, \textbf{k}$_T$, and \textbf{k}$_{3}$ can be precisely determined. Instead of as 3$\mathbf{k}_L$, 3$\mathbf{k}_T$ , or $\mathbf{k}_L$+2$\mathbf{k}_T$, the $\mathbf{k}_3$-type satellite peaks can only be indexed as 2$\mathbf{k}_L$+$\mathbf{k}_T$, as shown in Fig.\ref{fig:3rd}(c) (see also Fig.14 in the Appendix F), for the whole temperature range below $T_1$. The presence of such an inter-modulation high-order harmonics is usually seen as a strong evidence for the emergence of a multi-$\mathbf{k}$ magnetic structure \cite{Forgan1989,Adams2011,Marcus2018,Takagi2018,Puphal2020}. We thus conclude that Mn$_{3}$Sn possesses an intriguing double-$k$ non-coplanar magnetic structure below $T_{1}$, which can be regarded as a superposition of a longitudinal spin-density wave \textbf{k}$_L$ and a coplanar helical magnetic order \textbf{k}$_T$, both propagating along the $c$ axis. Such type of double-$k$ incommensurate magnetic order and the observation of the inter-modulation harmonics of the type of 2\textbf{k}$_L$+\textbf{k}$_T$ are extremely rare since \textbf{k}$_T$ and \textbf{k}$_L$ are not from the symmetric arms of the same modulation wavevector group or they occur on different magnetic elements. As shown in Fig.\ref{fig:3rd}(c), four phase regimes could be identified: \textbf{k} = 0 in the inverse triangular antiferromagnetic parent phase; \textbf{k}$_T$ \textless\ \textbf{k}$_L$ from 280 to 240 K; \textbf{k}$_T$ = \textbf{k}$_L$ in the lock-in phase from 240 to 200 K; and \textbf{k}$_T$ \textgreater\  \textbf{k}$_L$ below 200 K. Furthermore, clear "kinks" in the temperature dependence of the magnetic modulation wavevectors can be seen at 240 and 200 K, i.e., at the phase-boundary temperatures, reflecting a complex evolution of an apparent strong interaction between these two magnetic order parameters.\\

\subsection{Magnetic structure in the lock-in phase at 225 K}

While a complete determination of the double-$k$ non-coplanar incommensurate magnetic structures at the different phase regimes via neutron crystallographic approach can be a formidable task, the existence of this commensurate lock-in phase between 240 and 200 K with a single modulation wavevector \textbf{q}$_{lock-in}$ = (0,0,0.1) provides a feasible way to achieve this. We have thus investigated the magnetic structure at 225 K via single-crystal neutron diffraction at HEiDi. The magnetic structure refinement, based on in total 612 reflections (including 142 nuclear Bragg reflections and 470 magnetic satellites) measured at HEiDi, was performed based on the magnetic superspace approach via Jana2006 \cite{Perez-Mato2012,Peticek2010,Petricek2014}. Jana2006 also performs the representation analysis, and automatically generates all possible magnetic superspace symmetries for any given paramagnetic space group with only very basic input information, such as the crystal structure model and the propagation wavevector \cite{Petricek2014}. By trials and errors, we have found that the best fitted magnetic superspace model is $P6322.1'(00g)-h00s$ as shown in Fig.\ref{fig:mag_str}(a-b) with the refined $R(obs)$ = 5.89\% and $R_w(obs)$ = 11.71\%. In the magnetic superspace approach, the magnetic moment of the atom $v$ can be expressed by a Fourier series of the type,

\begin{equation}
\begin{aligned}
\mathbf{M}_{v}\left(\mathbf{k} \cdot \mathbf{r}_{v}\right)=& \mathbf{M}_{v, 0}+\sum_{m}\left[\mathbf{M}_{v, m \mathrm{s}} \sin \left(2 \pi m \mathbf{k} \cdot \mathbf{r}_{v}\right)\right.\\
&\left.+\mathbf{M}_{v, m \mathrm{c}} \cos \left(2 \pi m \mathbf{k} \cdot \mathbf{r}_{v}\right)\right]
\end{aligned}
\end{equation}

\noindent where $\mathbf{M}_{v, 0}$, $\mathbf{M}_{v, m \mathrm{s}}$ and $\mathbf{M}_{v, m \mathrm{c}}$ are the absolute value, the amplitude of the sine term and the amplitude of the cosine term, respectively. The details of the determined magnetic structure in the lock-in phase at 225 K are given in the Tab.\ref{tab:mag225_struture},

\begin{table}[htbp]
	\caption[]{\label{tab:mag225_struture}
		Refined magnetic structure parameters for the lock-in phase of Mn$_{3}$Sn at 225 K based on the magnetic superspace model $P6322.1'(00g)-h00s$.
	}
	\begin{center}
%		\begin{tabular}{lp{2cm}p{2cm}p{2cm}p{2cm}p{2cm}p{2cm}p{2cm}ll}
		\begin{tabular}{cp{0.8cm}p{0.8cm}p{1.3cm}p{1.3cm}p{1cm}p{1cm}p{1cm}cc}
			\hline
			\hline
			&Atom&Site&$x$&$y$&$z$&$Occ.$&$U_{iso}$ (\AA$^2$)&& \\
			\hline
			&Mn&$6h$&0.8383(2)&0.6766(2)&0.25&1.016&0.0050(3)&& \\
			&Sn&$8j$&0.3333&0.6667&0.25&0.3333&0.0040(5)&& \\
			\hline
			&\multicolumn{8}{l}{Refined magnetic components in the Bohr magneton:}&\\
			\hline
			&Waves&\multicolumn{8}{c}{amplitude ($\mu_{B}$)}\\
			\hline
			&0&\multicolumn{8}{c}{0.000(0)}\\
			&sin&\multicolumn{8}{c}{3.052(101)}\\
			&cos&\multicolumn{8}{c}{2.732(116)} \\
			\hline
			&\multicolumn{8}{l}{\textrm{$R(obs)$ = 3.43\%,  $R_{w}(obs)$ = 5.27\%,  $R_{w}(all)$ = 5.49\%}}&\\
			\hline
			&\multicolumn{8}{l}{$Occ.$: atomic occupation.}&\\
			&\multicolumn{8}{l}{$U_{iso}$ : atomic displacement parameters.}&\\
			\hline
			\hline
		\end{tabular}
	\end{center}
\end{table}

This magnetic structure model is indeed a non-coplanar magnetic order that can be decomposed into a longitudinal SDW along the $c$ direction (see Fig.\ref{fig:mag_str}(d)) and an elliptical helix in the $ab$ plane (as shown in Fig.\ref{fig:mag_str}(e)). In each kagome layer, the in-plane projection of magnetic moments retains the same inverse triangular magnetic configuration, but with a phase shift between layers. Possible topological Hall effects may arise in such a non-coplanar magnetic structure due to the presence of a finite scalar spin chirality, which is defined as $\kappa_{\mathrm{ijk}}=\boldsymbol{s}_{\mathrm{i}} \cdot (\boldsymbol{s}_{\mathrm{j}} \times \boldsymbol{s}_{\mathrm{k}})$, where $\boldsymbol{s}_{\mathrm{i}}$, $\boldsymbol{s}_{\mathrm{j}}$, $\boldsymbol{s}_{\mathrm{k}}$ are the non-coplanar spins within a triangle. However, as illustrated in Fig.\ref{fig:mag_str}(c), for the modulated magnetic structure in the lock-in phase there is a global time-reversal symmetry connecting the layer $l_{ci}$ and $l_{(c+5)i}$, thus leading to the vanishing of the total scalar spin chirality.\\

\begin{figure*}[htbp]
	\centering
	\includegraphics[width=14cm]{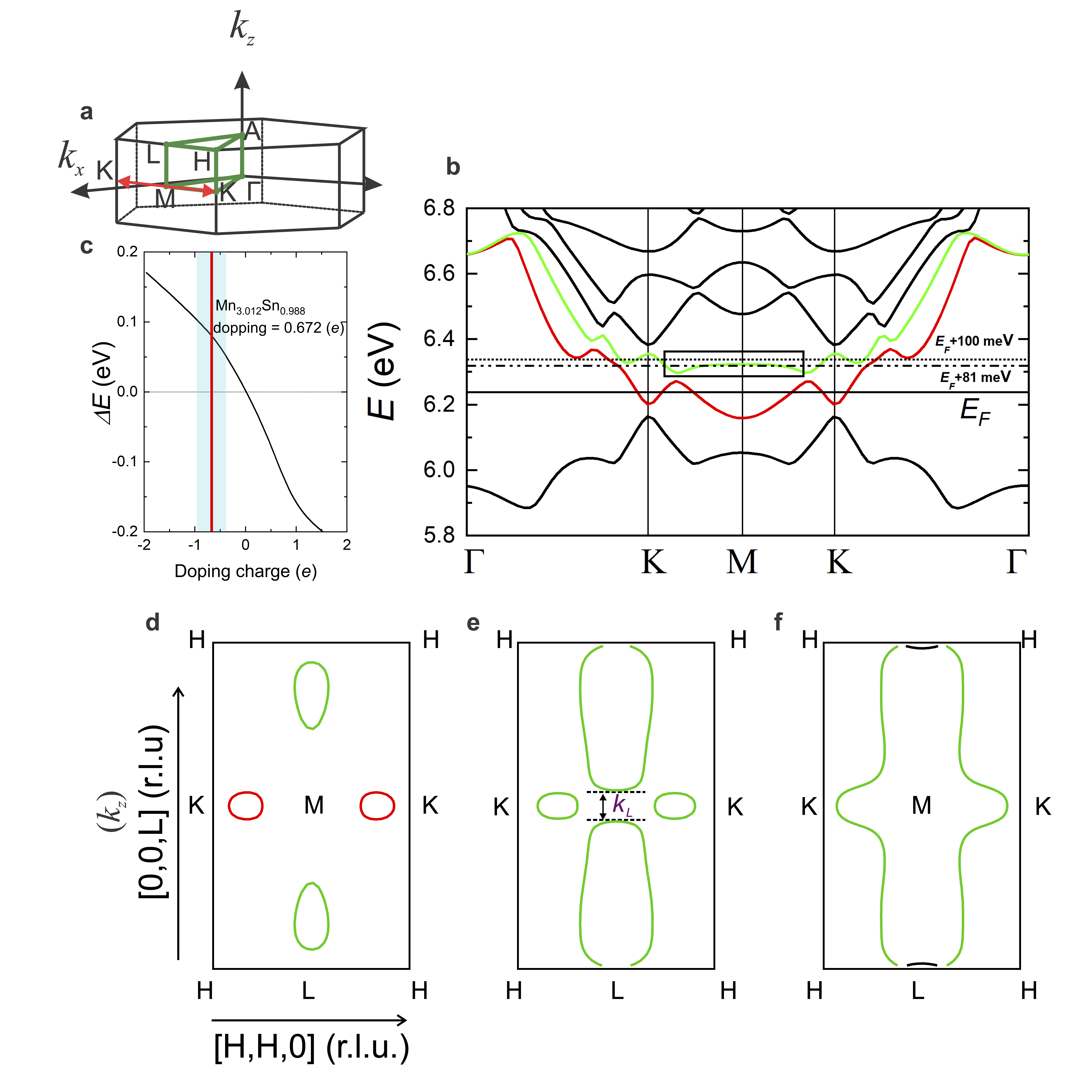}
	\caption{\label{fig:E_dispersion}
		Electronic band structure and the Fermi-surface nesting in Mn$_{3}$Sn from first-principles DFT calculations.
		\textbf{a}, The Brillouin zone of the hexagonal Mn$_{3}$Sn, where the reciprocal-space directions and high-symmetry points are marked.
		\textbf{b}, The band structure along the $\Gamma-K-M-K-\Gamma$ direction in Mn$_{3}$Sn that is calculated for the stoichiometric composition.
		\textbf{c}, The effects of the shift of the Fermi level $E_{F}$ as a function of the doping charge $e$. The effective doping charge based on the determined chemical composition of the studied sample is 0.672 $e$, which would shift $E_{F}$ up by about 81 meV. The light-blue region marks the likely charge doping range based on the refined error bars of the Mn occupancy in the studied sample.
		\textbf{d-f}, The illustration of the Fermi-surface nesting conditions in Mn$_{3}$Sn: for both (\textbf{d}) the stoichiometric (i.e., $E_{F}$+0 meV), and (\textbf{f}) the over-doped (i.e., $E_{F}$+100 meV) cases, there exists no Fermi-surface nesting condition; while for the weakly doped case, for instance, in our studied sample (i.e., $E_{F}$+81 meV), as shown in (\textbf{e}), the Fermi-surface nesting may happen. The occurrence of the Fermi-surface nesting is highly sensitive to the level of the doping charge, and the composition of the single crystals of Mn$_{3}$Sn grown from the Sn self-flux method is more likely in the weakly doped regime.
	}
\end{figure*}
  
\subsection{Electronic band structures, flat band and Fermi-surface nesting}

While the emergence of this longitudinal SDW from a magnetically ordered parent phase is rather peculiar, the clear jump in electric resistivity at $T_{1}$ (see Fig.\ref{fig:transition}(f)) would serve as a strong indication for the opening of an energy gap near $E_{F}$ in electronic band structures due to the formation of SDW, as often observed in other known SDW and CDW materials \cite{Fawcett1994,Wu2008}. We now turn to first-principles DFT band structure calculations for further insights. Our DFT calculations are based on the crystal structure and the room-temperature inverse triangular antiferromagnetic structure  that are experimentally determined via single-crystal neutron diffraction on the studied single crystal (see the Appendix A). While the calculated band structures are largely in resemblance to those reported in the previous work \cite{Kubler2014,Kubler2017,Kuroda2017,Yang2017}, a particularly interesting feature that is revealed in our calculations is the flat band (i.e. green-coloured) along the high-symmetry $K$-$M$-$K$ direction, as shown in Fig.\ref{fig:E_dispersion}(b). Given that as-grown Mn$_{3}$Sn is stable only with slightly excess Mn at the Sn sites, which would in turn act effectively as electron doping, we also analysed the chemical shift of $E_{F}$ as a function of charge doping, as shown in Fig.\ref{fig:E_dispersion}(c). The actual chemical composition of our studied crystal as determined from neutron diffraction structural refinement is Mn$_{3.012}$Sn$_{0.988}$ (see the Appendix A), equal to an effective electron doping of 0.672 $e$, so this doping level would shift $E_{F}$ up by about 81 meV. As shown in Fig.\ref{fig:E_dispersion}(b), due to this intrinsic chemical doping and the resulted shift of $E_{F}$, the \textit{K}-\textit{M}-\textit{K} flat band would reside exactly in the Fermi surface in the studied Mn$_{3}$Sn sample. This would create a favourable condition for the emergence of possible correlated many-body phenomena. Further analysis of the geometry of the Fermi surface for three representative cases of the doping level, namely, in the ideal stoichiometry (Fig.\ref{fig:E_dispersion}(d)), with optimal electron doping (Fig.\ref{fig:E_dispersion}(e)), and in the slightly over-doped case (Fig.\ref{fig:E_dispersion}(f)), has been undertaken. We found that, only with optimal electron doping in the case of Mn$_{3.012}$Sn$_{0.988}$, the parallel sections of this flat band in the Fermi surface could form a perfect nesting condition that matches to the modulation wavevectors of the observed incommensurate magnetic structures. We thus argue that the magnetic phase transition at $T_1$ is primarily driven by a SDW instability that is associated to the Fermi-surface nesting of flat bands, also a hallmark correlated electron phenomenon.\\

\subsection{Magnetic-field dependence of the multi-\textit{k} magnetic state below $T_1$}

\begin{figure}[htbp]
	\centering
	\includegraphics[width=8.5cm]{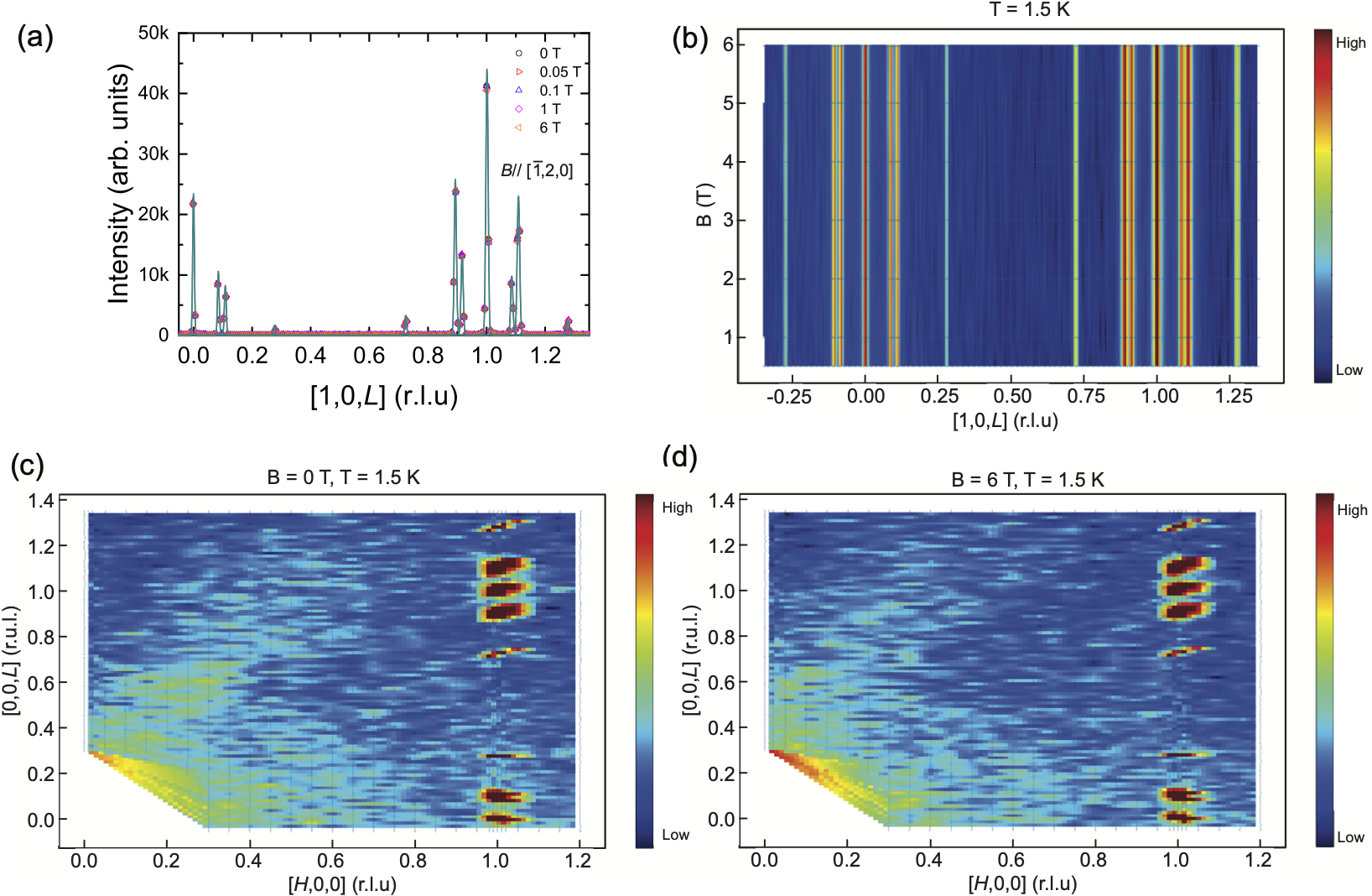}
	\caption{\label{fig:field_dependence}
		Field dependence of the magnetic satellite reflections in Mn$_3$Sn at 1.5 K measured at D23 with magnetic fields in parallel to the [$\overline{1}$,2,0] direction.
		\textbf{a-b}, $Q$-scan along the [1,0,\textit{L}] direction under various magnetic fields.
		\textbf{c-d}, Intensity mappings measured in the ($h$,0,$l$) reciprocal-space plane under 0 T and 6 T, respectively.
	}
\end{figure}

The magnetic-field dependent measurements of this emergent multi-\textit{k} magnetic state were carried out at D23. For this experiment, the ($h$,0,$l$) reciprocal plane of the studied crystal is aligned in the horizontal scattering plane, so that the applied magnetic field is in parallel to the [$\overline{1}$,2,0] direction of this hexagonal lattice system. As shown in Fig.\ref{fig:field_dependence}(a-b), no changes could be seen from the field dependence of the $Q$-scan along the [1,0,\textit{L}] direction in the range of 0 to 6 T at 1.5 K. This indicates that the low-temperature non-coplanar double-$k$ incommensurate phase is robust against magnetic fields applied along the [1,0,\textit{L}] direction. Furthermore, comprehensive mappings in the ($h$,0,$l$) reciprocal-space plane as a function of applied magnetic fields were measured, as shown in  Fig.\ref{fig:field_dependence}(c-d). No evidence for possible new field-induced long-range or short-range magnetic modulations, such as an emergent magnetic skyrmion lattice suggested in a recent powder based study \cite{Rout2019}, could be seen from these reciprocal-space mappings in our studied single-crystal sample. Therefore, the observed topological Hall effect (THE) in the mentioned powder sample \cite{Rout2019} is not necessarily related to a well formed non-coplanar spin textures such as magnetic skyrmions, but possibly due to extrinsic causes, such as the chemical composition inhomogeneity, off-stoichiometry and multiple magnetic domains etc. in powders. It would be interesting to carry out further field-dependent investigations under applied fields in parallel to the [0,0,1] direction on the flux-method grown crystals.\\

\section{Discussion and summary}

At first thought, the emergence of SDW from an antiferromagnetically ordered parent phase may seem very strange. But, based on our DFT band-structure calculations, the coplanar and non-collinear magnetic order in the parent phase is actually needed as one necessary condition for the formation of the identified topological flat bands and the resultant Fermi-surface nesting between them (also see the Appendix G). Another two necessary conditions are the chemical potential shift due to excess Mn, which can tune the Fermi level close to the flat band, and the electronic correlation effects, which would further sharpen the density of states (DOS) of the flat bands and thus favour emergent many-body phenomena. Such a correlated electronic band structure in Mn$_3$Sn is indeed captured in a recent DFT+DMFT study \cite{Yu2022}. The formation of SDW would in turn open an energy gap at $E_{F}$ and thus considerably alter the electronic structure near $E_{F}$. For instance, given that the same inverse triangular spin configuration is largely retained in the individual kagome layer of Mn below $T_1$, the complete suppression of AHE must be related to a dramatic modification of Berry curvature of the relevant electronic bands due to this significant band renormalisation. Furthermore, a potential impact on the Weyl fermions from SDW as well as a possible interplay between them are expected, which could become a fascinating topic to explore in the future.\\ 

Unlike a usual helical order with a constant magnetic moment and phase between spins related by a lattice translation along the propagation vector, the helix in the realised double-$k$ non-coplanar magnetic phase is incommensurately modulated with the periodicity of an amplitude-modulated SDW. The ground-state magnetic structure in Mn$_3$Sn thus decomposes into a set of magnetic order parameters with distinct propagation vectors. The very presence of the inter-modulation high-order harmonics, as well as the intriguing temperature dependence of these distinct propagation vectors (e.g., clear "kinks" in the phase boundary as shown in Fig.\ref{fig:3rd}(c)), clearly indicate that this exotic magnetic phase is a result of the strong coupling between the emergent SDW and the high-temperature magnetic parent phase with an inverse triangular spin configuration. We believe that this coupling is achieved mainly via a very subtle effect of the amplitude-modulated SDW on the inter-unit cell exchange interaction term $J_6$ \cite{Park2018} (see Fig.\ref{fig:mag_str}(a)). In fact, in a similar way, a coupling between a multi-\textit{k} helical order and orbital-density wave was also realised in a different helimagnet \cite{Johnson2016}. Furthermore, as shown in Fig.\ref{fig:field_dependence}, since no changes could be found from the magnetic-field dependent measurement in the range of 0-6 T at 1.5 K, this suggests that the low-temperature double-$k$ non-coplanar incommensurate phase is robust against magnetic fields. This also strongly supports our interpretation that the phase transition at $T_1$ is primarily driven by emergent many-body electronic phenomena, such as SDW, since they act in larger energy scales than the antisymmetric DMI interaction. In this regard, this highly unusual double-$k$ non-coplanar incommensurate magnetic state is simply a consequence of the competition and coexistence between the electronic correlation-driven SDW and DMI-driven no-collinear chiral magnetism.\\

Our discovery also represents an extraordinary example for an intrinsic flat-band engineering of electronic and magnetic properties in topological metals. While the shifting of the Fermi level is achieved via intrinsic doping from excess Mn in the studied samples, it can be expected that this may also be realised via the controlled extrinsic chemical substitutions, the applied pressure or uniaxial strain in a similar manner. The effect of the applied uniaxial strain on the SDW phase transition in Mn$_3$Sn has indeed been found recently \cite{Ikhlas2020a,Singh2020,Ikhlas2022}. Furthermore, the tuning of the Fermi level via systematic chemical substitutions, as well as its impact on the interplay between CDW and superconductivity, were also demonstrated in the kagome superconductor CsV$_3$Sb$_5$ \cite{Oey2022, Li2022}. The proximity of a flat band-engineered SDW instability to a room-temperature no-collinear antiferromagnetic order can lead to even more exotic properties and functionalities that are manipulatable for this interesting topological quantum materials. \\

In summary, we have carried out a comprehensive investigation of the magnetic structures in the topological kagome metal Mn$_3$Sn via combined polarised and non-polarised single-crystal neutron diffraction. In consistence with the previous reports, a coplanar and non-collinear \textbf{k}=0 inverse triangular antiferromagnetic structure can be confirmed below $T_N$ = 420 K in our studied single-crystal samples. An additional phase transition that is accompanied by a substantial change simultaneously in electric transport, AHE and magnetic susceptibility is identified at $T_{1}$ = 280 K. Based on neutron polarisation analysis and the magnetic structure refinement, it is revealed that Mn$_3$Sn transforms to an intriguing double-$k$ non-coplanar incommensurate magnetic structure below $T_{1}$. The double-$k$ nature of this complex low-temperature magnetic order is unambiguously confirmed by the observation of the inter-modulation high-order harmonics of the type of 2\textbf{k}$_L$+\textbf{k}$_T$. Furthermore, our DFT band-structure calculations quantitatively demonstrate that the small electron doping due to the intrinsic Mn excess is responsible for the shift of $E_{F}$, so that this $K$-$M$-$K$ flat band can be intrinsically engineered to be exactly at the Fermi surface for the studied samples, and this thus paves the way for the emergence of rich Fermi surface-mediated electron correlation effects. We further argue that this intriguing magnetic phase transition at $T_{1}$ is primarily driven by a longitudinal SDW instability that is associated to the Fermi-surface nesting between the identified flat bands. The discovery of flat band-engineered SDW in Mn$_{3}$Sn not only solves a long-standing puzzle concerning the nature of the phase transition at $T_{1}$, but also provides an extraordinary example on the interplay and possible intrinsic manipulation of topology and electron correlation in topological matter. The identified multi-\textit{k} magnetic states in such by now prototypical example of magnetic topological materials can have repercussions for exploring novel topological phases in magnetic materials, and may offer new possibilities for the engineering of the novel modes of magnetization and chirality switching in Mn$_{3}$Sn and the related materials for potential applications in topological and antiferromagnetic spintronics.\\

\section*{Acknowledgments}

X.W. acknowledges V. Petř{\'{i}}{\v{c}}ek for the fruitful discussions on the single-crystal structure refinement with Jana2006, and J. M. Perez-Mato for the help on the use of the Bilbao Crystallographic Server. X.W. acknowledges the financial support from the China Scholarship Council. F.Z. and J.S. acknowledge the funding supports from the HGF–OCPC Postdoctoral Program. We acknowledge S. Mayr for assistance with the crystal orientation, and T. Herrmannsdoerfer and S. Chattopadhyay for their help in the preliminary characterisations of our single-crystal samples. Y.S. acknowledges T. Br\"uckel, J. K\"ubler and D. Khalyavin for valuable discussions. This work is based on the neutron diffraction experiments performed at DNS (MLZ, Garching), HEiDi (MLZ, Garching), IN12 (ILL, Grenoble). D23 (ILL, Grenoble) and FLEXX (HZB, Berlin) neutron instruments. W.F. acknowledges the National Key R\&D Program of China (Grant Nos. 2022YFA1403800 and 2022YFA1402600) and the National Natural Science Foundation of China (Grant Nos. 12274027 and 11874085). Y.M. and W.F. acknowledge the funding under the Joint Sino-German Research Projects (Chinese Grant No. 12061131002 and German Grant No. 1731/10-1) and the Sino-German Mobility Programme (Grant No. M-0142). Y.M. gratefully acknowledges financial support by the Deutsche Forschungsgemeinschaft (DFG, German Research Foundation) - TRR 288 - 422213477 (project B06). M.H. acknowledges the National Natural Science Foundation of China (Grant No. 11904040) and the Chongqing Research Program of Basic Research and Frontier Technology, China (Grant No. cstc2020jcyj-msxmX0263).\\

The authors declare that they have no competing interests.\\

Y.S. conceived and supervised the project.
W.F. supervised the work on DFT calculations.
X.W. carried out the single-crystal growth and characterisation of physical properties with the supports from C.Y. and Yo.S.
X.M. and M.H. carried out the electric transport measurements.
X.W., F.Z. and Y.S. carried out the neutron diffraction experiments with the supports from M.M., J.S., T.M., W.S., K.S., E.R. and J.X.
X.Y. and W.F. carried out the DFT calculations with the supports from Y.M. and S.B.
X.W. and Y.S. analysed and interpreted the results with the supports from Z.F., W.F., G.R.
X.W. and Y.S. wrote the manuscript with the contributions from all authors.
Correspondence and requests for materials should be addressed to Y.S. or W.F.\\

%\section{Appendix}

\section*{Appendix A: Details about the chemical composition analysis of the as-grown Mn$_{3}$Sn single crystals}

The chemical composition of the as-grown Mn$_{3}$Sn single crystals was determined by a combined study of energy dispersive X-ray analysis (EDX) and single-crystal neutron diffraction. The EDX results show that the average chemical composition of our as-grown crystals is Mn$_{3}$Sn$_{1.033}$, which is close to the stoichiometric composition. The slightly excess Sn is due to the residual Sn flux on the sample surface. Below $T_1$ = 280 K, as demonstrated in our polarised neutron diffraction experiments, the room-temperature non-collinear $\mathbf{k}$ = 0 antiferromagnetic order is completely disappeared in our studied samples. In addition, we have shown that, in the lock-in phase in the temperature range of 200-240 K, the magnetic satellite reflections are merged together, and meanwhile, are well separated from Bragg reflections. Therefore, we have collected 707 nuclear Bragg reflections at 210 K at HEiDi \cite{Meven2015} for the determinations of the crystalline structure and chemical composition of the as-grown Mn$_{3}$Sn single crystals. 

\begin{table}[htbp]
	\caption[Mn$_{3}$Sn nuclear refinement results]{\label{tab:struture}
		The determined crystalline structure of the as-grown Mn$_{3}$Sn sample based on the single-crystal neutron diffraction data taken at HEiDi at 210 K.
	}
	\begin{center}
%		\begin{tabular}{cp{2cm}p{2cm}p{2.5cm}p{2.5cm}p{2.5cm}p{2.5cm}p{2.5cm}c}
		\begin{tabular}{cp{0.8cm}p{0.8cm}p{1.5cm}p{1.7cm}p{0.7cm}p{1.1cm}p{1cm}c}
			\hline
			\hline
					&Atom&Site&$x$&$y$&$z$&$Occ.$&$U_{iso}$({\AA}$^{2}$)& \\
			\hline
			&Mn&$6h$&0.83865(8)&0.67729(16)&0.25&1.016(7)&0.0073(2)& \\
			&Sn&$2c$&0.3333&0.6667&0.25&0.3333&0.0063(3)& \\
			\hline
			&\multicolumn{8}{l}{\textrm{$a$= 5.67 \AA,  $c$ = 4.53 \AA}}\\
				\hline
			&\multicolumn{8}{l}{$Occ.$: atomic occupation.}\\
			&\multicolumn{8}{l}{$U_{iso}$: isotropic atomic displacement parameters.}\\
			&\multicolumn{8}{l}{\textrm{$R(obs)$ = 2.30\%,  $R_{w}(obs)$ = 2.59\%,  $R_{w}(all)$ = 2.64\%}}\\
			\hline
			\hline
		\end{tabular}
	\end{center}
\end{table}

\begin{figure}[htbp]
	\centering
	\includegraphics[width=6cm]{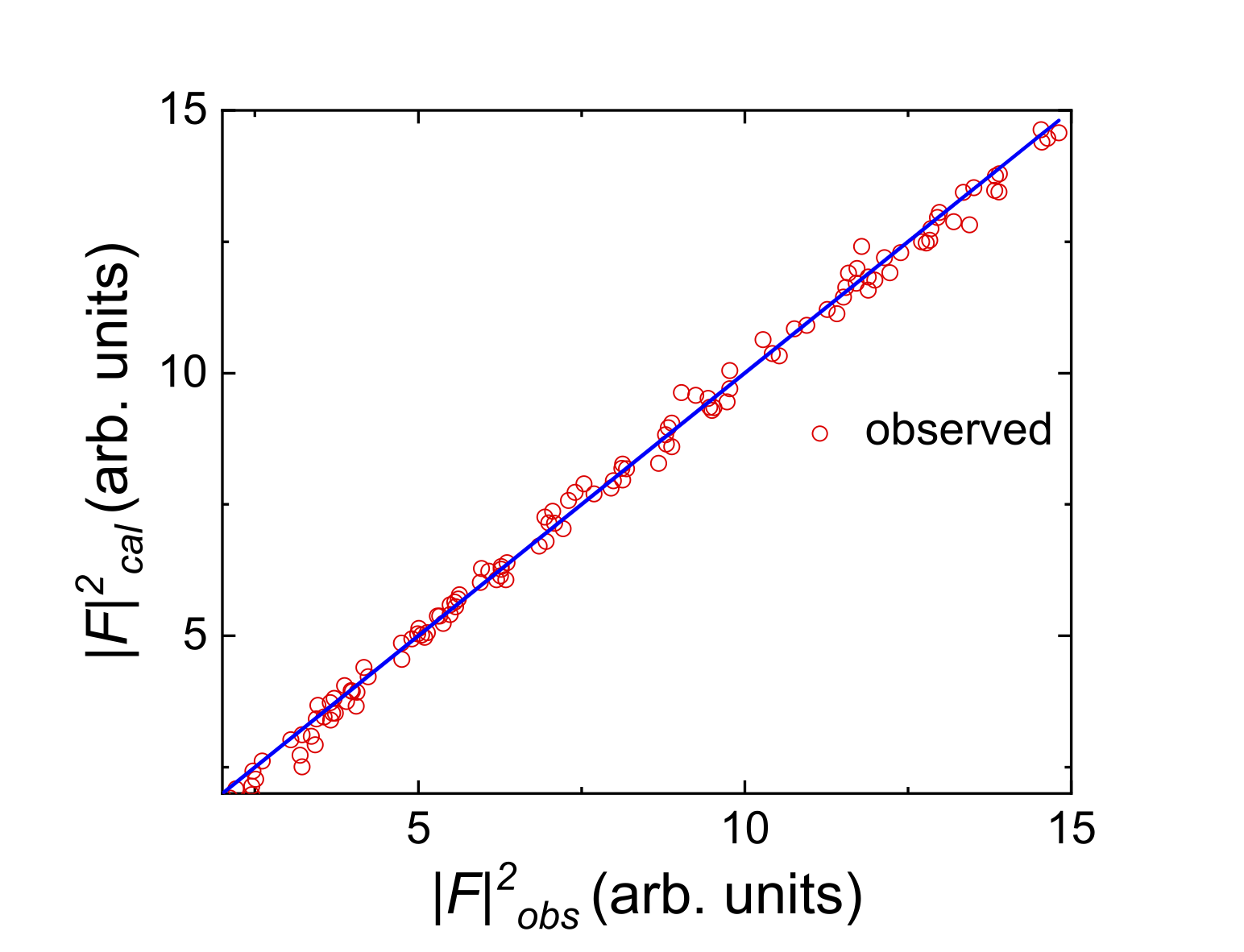}
	\caption{\label{fig:nuclear_210K}
		The calculated squared structure factors versus the measured ones, from the Jana2006 refinement based on the single-crystal neutron diffraction data taken at HEiDi at 210 K.
	}
\end{figure}

The structure refinement was performed with the crystallographic program Jana2006, and the refinement result is listed in Tab.\ref{tab:struture}. The refined weighted $R_{w}$, at 2.64\%, is very good. The refined chemical composition is Mn$_{3.012}$Sn$_{0.988}$, which is close to the stoichiometric composition and also clearly confirms the slight excess of Mn. Since no structural phase transition at low temperatures is found for this compound, we also use the nuclear structure obtained at 210 K as the input for the structural refinement at other temperatures. The calculated squared structure factors from the Jana2006 refinement versus the measured ones are shown in Fig.\ref{fig:nuclear_210K}. It is well known that Mn and Sn have opposite sign of neutron scattering length, which gives a large contrast and makes the precise determination of the site-specific composition by neutron diffraction possible.

\section*{Appendix B: Details about the flipping-ratio correction of the polarised neutron diffraction data}

\begin{figure}[htbp]
	\centering
	\includegraphics[width=7cm]{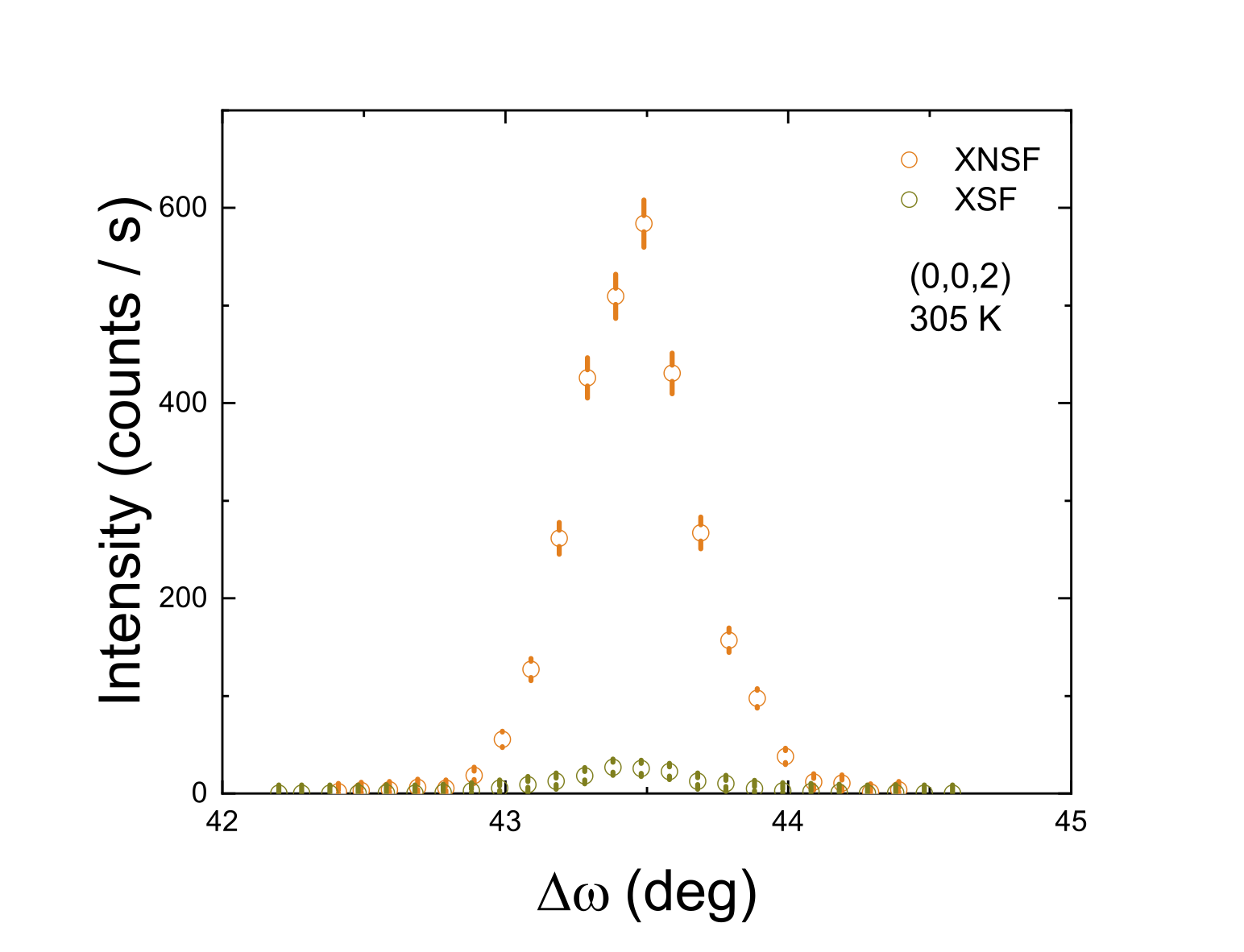}
	\caption{\label{fig:Ref_002_XX}
		Rocking-curve scans of the (0,0,2) reflection measured at 305 K in the $xsf$ and $xnsf$ channel, respectively.
		$\Delta$$\omega$ represents the relative angle of the sample rotation.
		For both the room-temperature magnetic structure models, \textit{Cm'cm'} or \textit{Cmc'm'}, the (0,0,2) reflection is a pure nuclear reflection. 
		Based on this peak, we can derive the flipping ratio of the employed experimental conditions.
	}
\end{figure}

The flipping-ratio correction was made for the polarised neutron diffraction data taken at IN12. For both room-temperature magnetic structure models, \textit{Cm'cm'} or \textit{Cmc'm'}, the (0,0,2) reflection is a pure nuclear reflection that has no magnetic contributions. Therefore, it is possible to check the flipping ratio by measuring the intensities of the $xnsf$ and $xsf$ channels of this reflection, which would reflect the polarisation rate of the incident polarised neutrons and the efficiency of polarisation analysers in the employed experimental condition. The measured flipping ratio $R = \frac{I_{nsf}-I_{nsf{\_}bg}}{I_{sf}-I_{sf{\_}bg}}$ for (0,0,2) at 305 K is about 20 as can be derived from Fig.\ref{fig:Ref_002_XX}, which is comparable to the flipping ratio measured on a reference sample of graphite. This indicates that there is no neutron depolarisation effect in the studied sample at room temperature. Note that $I_{nsf{\_}bg}$ and $I_{sf{\_}bg}$ are the respective background intensities in the the $nsf$ and $sf$ channels. By applying the flipping ratio of the (0,0,2) reflection, we could correct the polarised neutron diffraction data in a wide temperature range. The flipping-ratio correction can be performed via the following equations for each polarisation channel,

\begin{eqnarray}
&I_{i}^{corr{\_}n s f}=I_{i}^{n s f}+\frac{1}{R_{i}-1}\left(I_{i}^{n s f}-I_{i}^{s f}\right)\\
&I_{i}^{corr{\_}s f}=I_{i}^{s f}-\frac{1}{R_{i}-1}\left(I_{i}^{n s f}-I_{i}^{s f}\right)
\end{eqnarray}

where $i=x,y,z$ and $R_{i}$ is the measured flipping ratio at the corresponding polarisation channel.

\section*{Appendix C: Temperature dependence of the (1,1,0) reflection}

\begin{figure}[htbp]
	\centering
	\includegraphics[width=7cm]{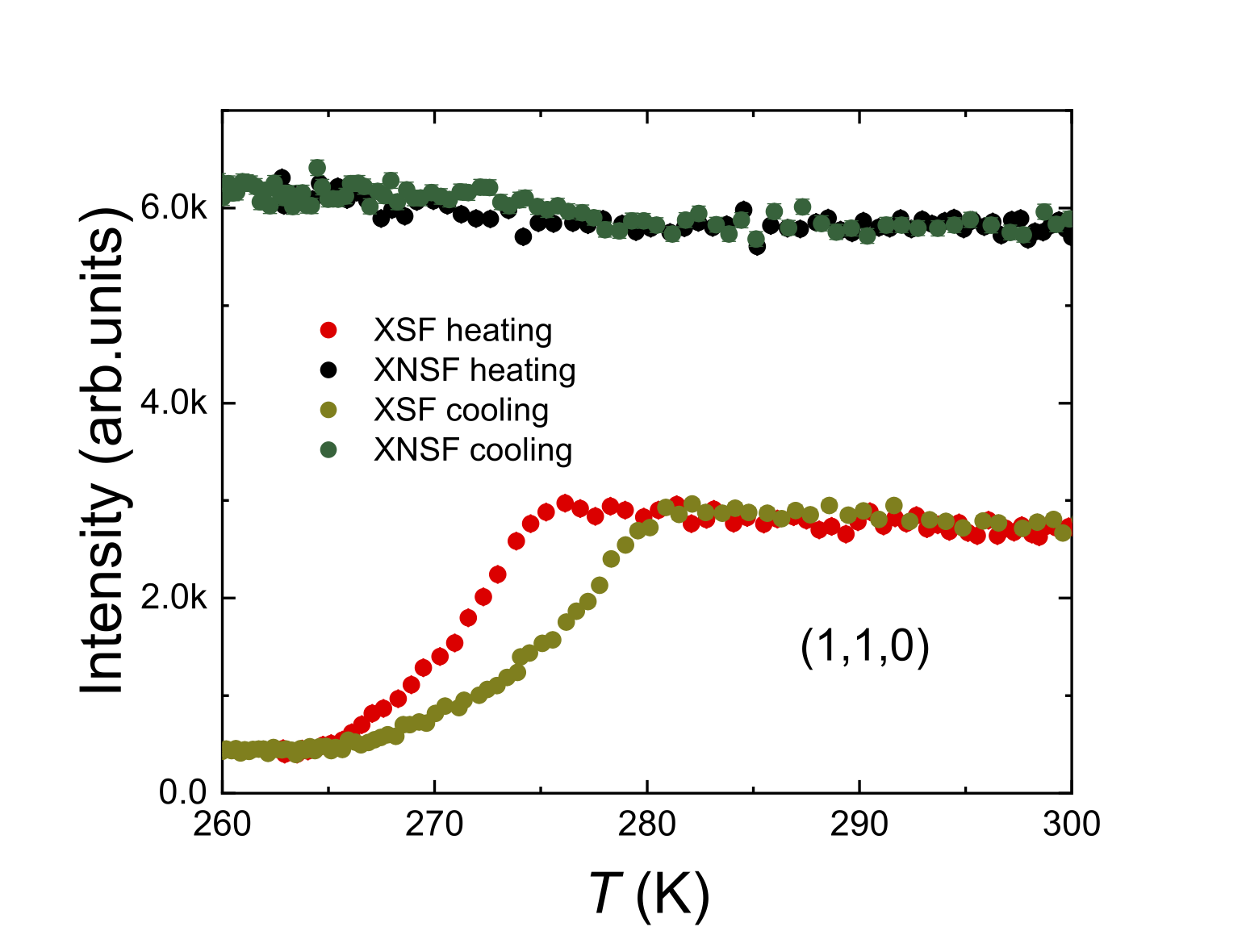}
	\caption{\label{fig:Ref_110_T_dep}
		Temperature dependence of the (1,1,0) refection measured in the $xsf$ and $xnsf$ channels upon heating and cooling.
	}
\end{figure}

The polarised neutron diffraction was also carried out on the single crystals that are ($h$,$h$,$l$) oriented in the horizontal scattering plane. As shown in Fig.\ref{fig:Ref_110_T_dep}, a magnetic phase transition was observed around $T_{1}$ from the the temperature dependence of the (1,1,0) reflection in the $xsf$ and $xnsf$ channels, consistent with the previous results. There is a small hysteresis upon heating and cooling, which is likely due to a combined effect from the different temperature changing rates and a slight composition inhomogeneity in the measured sample, but not an indication of a possible structural transition, as our neutron diffraction study has shown that the structure of Mn$_3$Sn at 210 K is same as that at room temperature.
 
\section*{Appendix D: Sample dependence: Comparison of the phase transition at $T_1$ between our samples and a Bridgman-method grown crystal}

\begin{figure}[htbp]
	\centering
	\includegraphics[width=8cm]{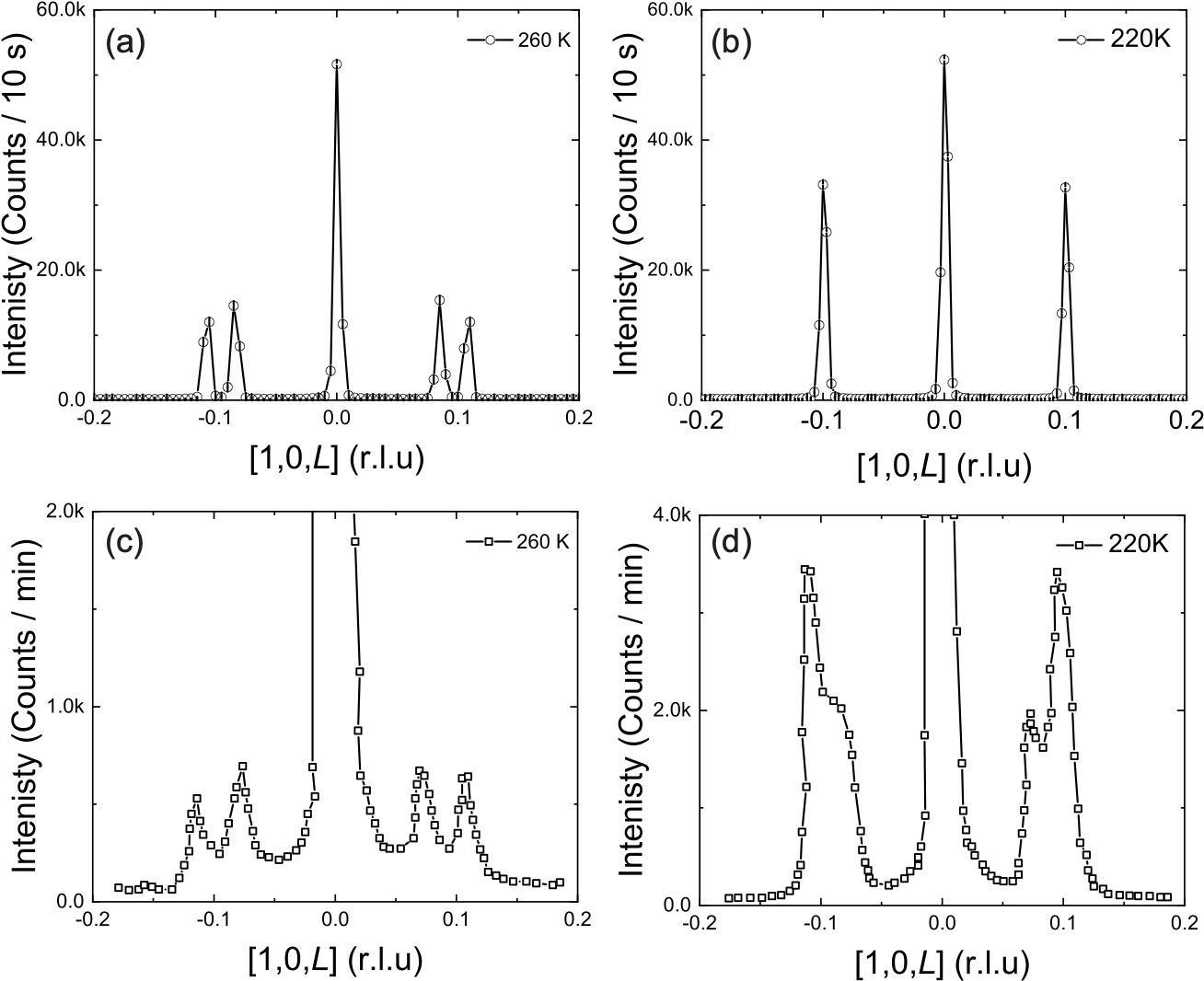}
	\caption{\label{fig:sample_comp}
		Comparison of the same $Q$-scans along the [1,0,$L$] direction between our flux-method grown sample and a Bridgman-method grown crystal taken from an earlier report \cite{Cable1993}. 
		(a) and (b): Scans measured in our flux-method grown sample at 260 K and 220 K, respectively.
		(c) and (d): Scans measured in the Bridgman-method grown crystal at 260 K and 220 K, respectively \cite{Cable1993}.
		 The (1,0,0) refection in (c) and (d) has a peak intensity of $\sim$ 71000 counts/minute.
	}
\end{figure}

There exists an apparent sample dependence for the phase transition at $T_1$ among single crystals grown by different methods \cite{Ikhlas2020,Sung2018}. A detailed comparison of the same $Q$-scans along the [1,0,\textit{L}] direction between our flux-method grown samples and a Bridgman-method grown crystal \cite{Cable1993} is illustrated in Fig.\ref{fig:sample_comp}. It can be clearly noticed that at 260 K, the intensity of the (1,0,0) reflection in our studied sample is about 51000 counts/10s, and that of the magnetic satellite reflections in the incommensurate phase is almost 14000 counts/10s, leading to an intensity ratio of about 3.64. In comparison, for the Bridgman-method grown crystal, the intensities for the (1,0,0) reflection and the corresponding magnetic satellites are 71000 and 700 counts/minute, respectively, thus giving rise to a ratio of 101.43 at 260 K. The extreme weakness of these magnetic satellite reflections is a clear evidence that the phase transition at $T_1$ in this Bridgman-method grown crystal only occurs in a very small volume fraction and that the significantly large portion of the crystal still remains in the high-temperature $\mathbf{k}$ = 0 inverse triangular antiferromagnetic order phase. In contrast, in our careful polarised neutron diffraction experiments, the magnetic components of the (1,0,0) and (1,1,0) reflections, which are characteristic to the high-temperature $\mathbf{k}$ = 0 phase, are seen completely disappeared below $T_1$, thus unambiguously demonstrating that the phase transition at $T_1$ in our samples is thorough and complete. Furthermore, in the lock-in phase at 220 K in our sample, the two magnetic satellites are merged together into a single sharp reflection that has a comparable intensity to that of (1,0,0). However, in the Bridgman-method grown crystal two magnetic satellites do not completely coincide with each other at 220 K, which is due to likely inhomogeneity in its chemical composition.
  
\begin{figure}[htbp]
	\centering
	\includegraphics[width=8cm]{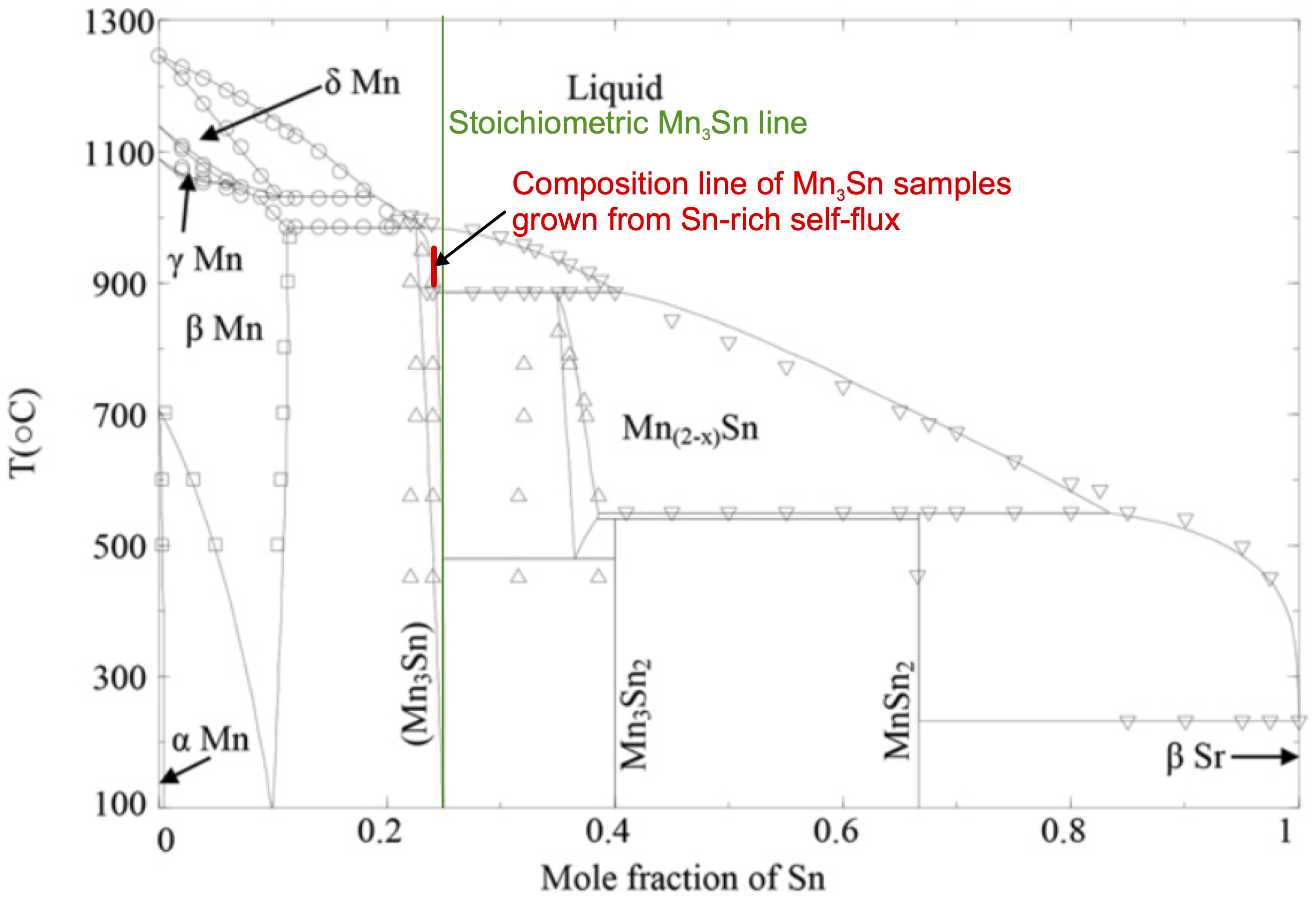}
	\caption{\label{fig:phaseD}
		A Mn-Sn binary phase diagram taken from the ref. \cite{Aljarrah2015}, in which both the stoichiometric (green coloured) and actual composition (red coloured) lines are drawn for comparison. 
	}
\end{figure}

The sample dependence of physical properties among Mn$_{3}$Sn crystals grown with different preparation methods can be explained based on the Mn-Sn binary phase diagram as shown in Fig.\ref{fig:phaseD} \cite{Aljarrah2015}. It can be seen that the phase of Mn$_{3}$Sn is stabilised only when the excess Mn is present. For a crystal grown from the Sn-rich self-flux, its chemical composition should be very close to the stoichiometric one so that the intrinsic Mn excess is minimised in the sample. Therefore, the phase diagram can ensure that the samples grown from the Sn flux method even by different groups still have the similar chemical compositions, and would more likely show the phase transition at $T_1$ \cite{Sung2018,Yan2019}. However, for the crystals grown from the Bridgman method, their actual chemical compositions are highly dependent on the compositions of the starting materials, the growth speed and the cooling rate. A disadvantage of the flux method for the growth of Mn$_{3}$Sn is the very narrow growth window in 884-984$^\circ$C with the risk of the precipitation of the ferromagnetic Mn$_{2}$Sn. Strictly speaking, Mn$_3$Sn does not melt congruently, so the Bridgman method may not be the method of choice if the exact stoichiometry is important.

\section*{Appendix E: Neutron polarisation analysis results of the magnetic satellite reflections (1,0,1-$\delta$) at 180 K}

\begin{figure}[htbp]
	\centering
	\includegraphics[width=7cm]{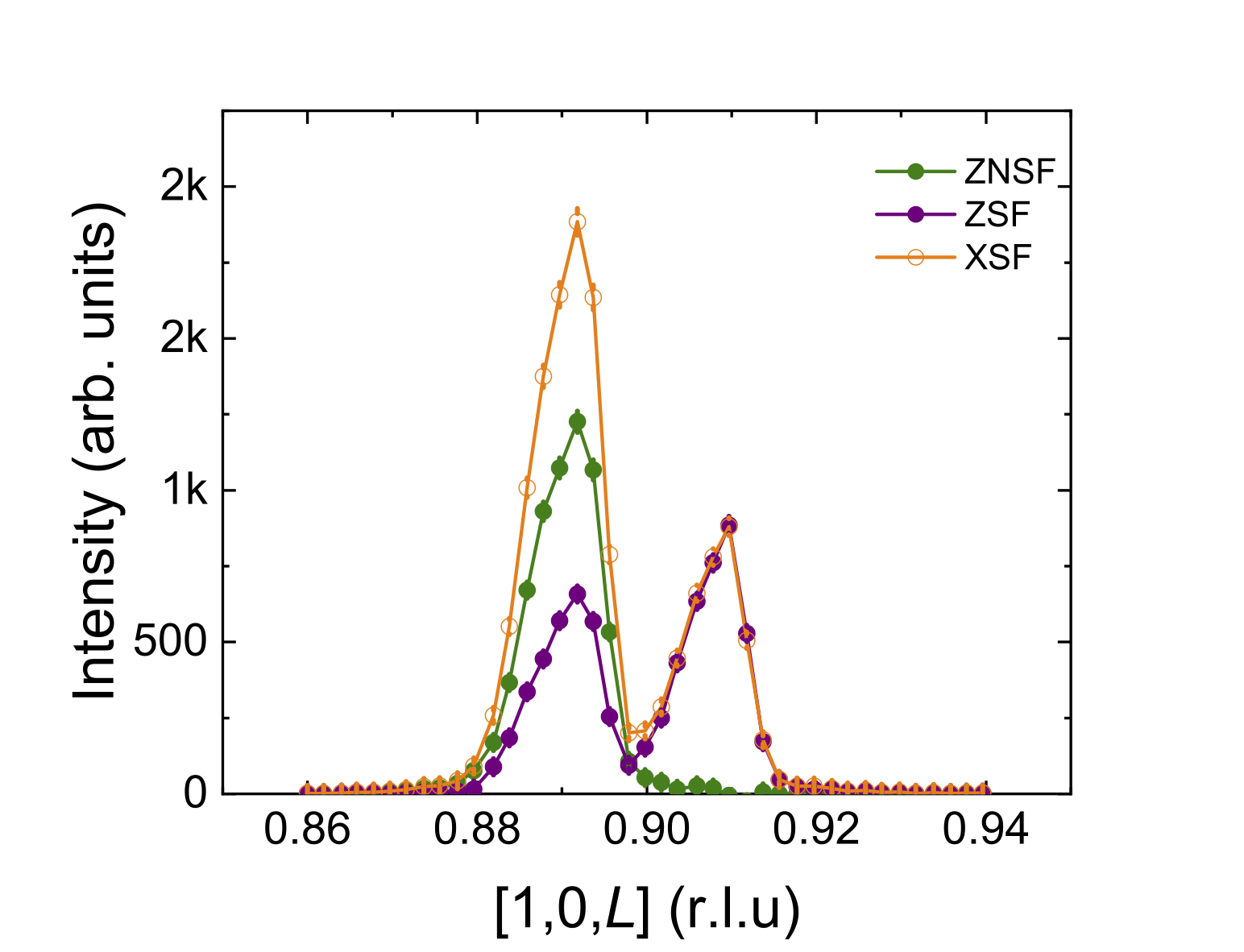}
	\caption{\label{fig:Ref_101_180K}
		$Q$-scans of the magnetic satellite reflections of the (1,0,1-$\delta$) type at 180 K measured in the $xsf$, $zsf$ and $znsf$ channels, respectively.
	}
\end{figure}

Neutron polarisation analysis is also carried out for the magnetic satellite reflections of the (1,0,1-$\delta$) type in the $xsf$, $zsf$ and $znsf$ channels in the double-$k$ modulated incommensurate phase at 180 K, as shown in in Fig.\ref{fig:Ref_101_180K}. It is anticipated that for these magnetic satellite reflections double peaks would appear in the $zsf$ channel and a single peak in the $znsf$ channel. This is because the angle $\theta_3$ between the scattering vector $Q_{3}$ = (1,0,$\sim$0.9) and the [1,0,0] direction is about 45$^\circ$, and we will have the following equations:

\begin{equation}
\begin{array}{l}
\sigma_{z}^{zsf}=\mathbf{M}_{\perp y}^{2}\\=|\mathbf{M}^{[1,0,0]}|^{2}\sin^{2}\theta_3
+ |\mathbf{M}^{[0,0,1]}|^{2}\cos^{2}\theta_3\\ = \frac{1}{2}(|\mathbf{M}^{[1,0,0]}|^{2}+|\mathbf{M}^{[0,0,1]}|^{2})\sigma_{z}^{znsf}
\\=\mathbf{M}_{\perp z}^{2}=|\mathbf{M}^{[\overline{1},2,0]}|^{2}
\end{array}
\end{equation}

In this case, both the magnetic components in the $ab$ plane and along the $c$ direction contribute to the $zsf$ channel, and in the $znsf$ channel there will always be a single peak since the $z$ direction is perpendicular to the scattering plane, which is parallel to the [$\overline{1}$,2,0] direction in our scattering geometry.

\section*{Appendix F: Indexing of the inter-modulation high-order harmonic}

\begin{figure}[htbp]
	\centering
	\includegraphics[width=8cm]{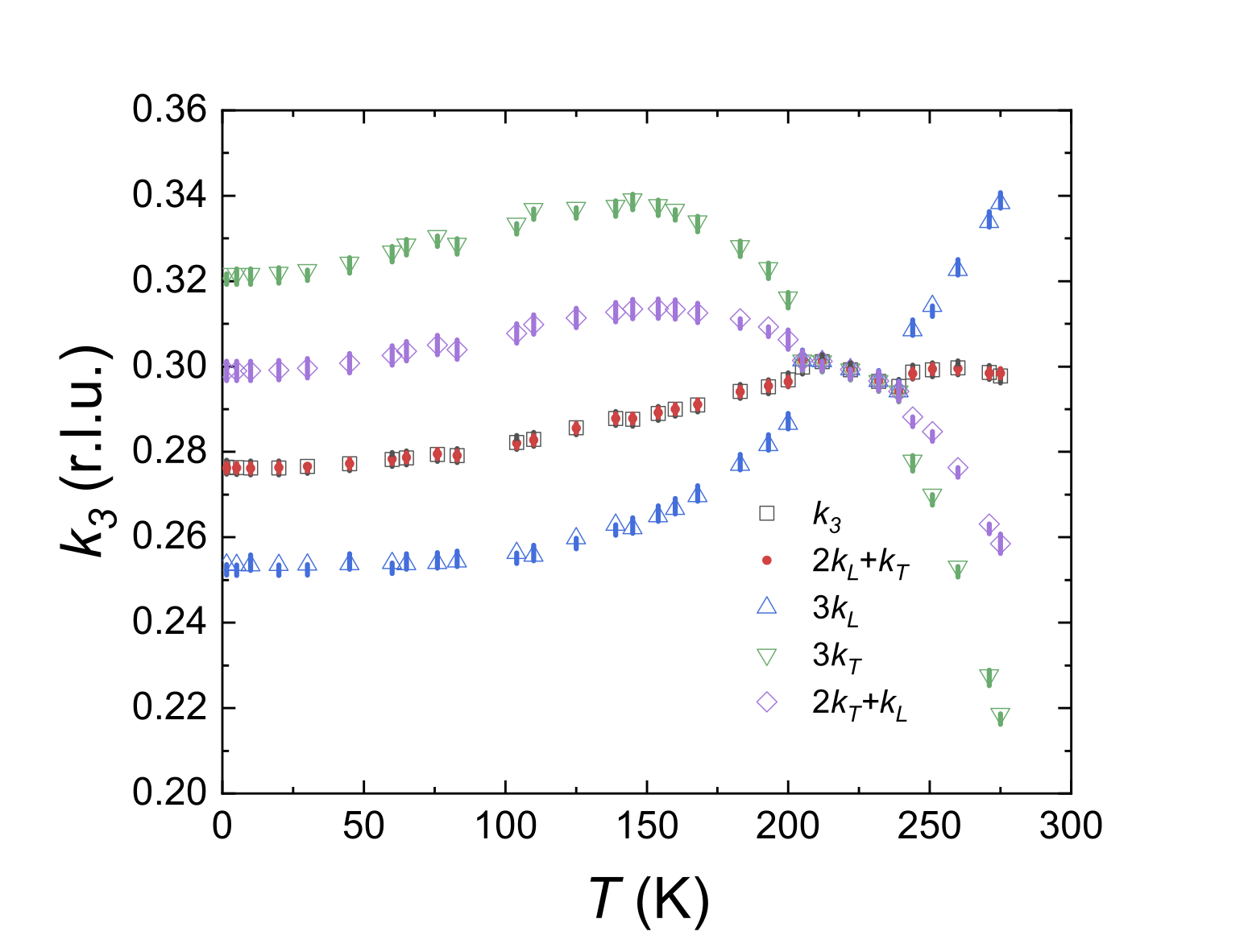}
	\caption{\label{fig:3rd_all}
		Indexing of the high-order harmonic $\mathbf{k}_{3}$.
	}
\end{figure}

The modulation wavevector of the high-order harmonic $\mathbf{k}_{3}$ is obtained over a wide temperature range from the high-resolution $Q$-scans measured at D23. In the previous reports \cite{Cable1993}, $\mathbf{k}_{3}$ was indexed as the third order of $\mathbf{k}_{T}$ (i.e., $3\mathbf{k}_{T}$). Our high-resolution measurements clearly indicate that this is incorrect. As shown in Fig.\ref{fig:3rd_all}, among various possibilities, the high-order harmonic $\mathbf{k}_{3}$ could only be indexed as
the inter-modulation high-order harmonics of the type of $2\mathbf{k}_{L}+\mathbf{k}_{T}$. Such kind of inter-modulation high-order harmonics is a characteristic feature of a multiple-$\mathit{k}$ magnetic structure.
 
\section*{Appendix G: Detailed electronic band structures of Mn$_{3}$Sn obtained from the DFT calculations }

\begin{figure}[htbp]
	\centering
	\includegraphics[width=8.5cm]{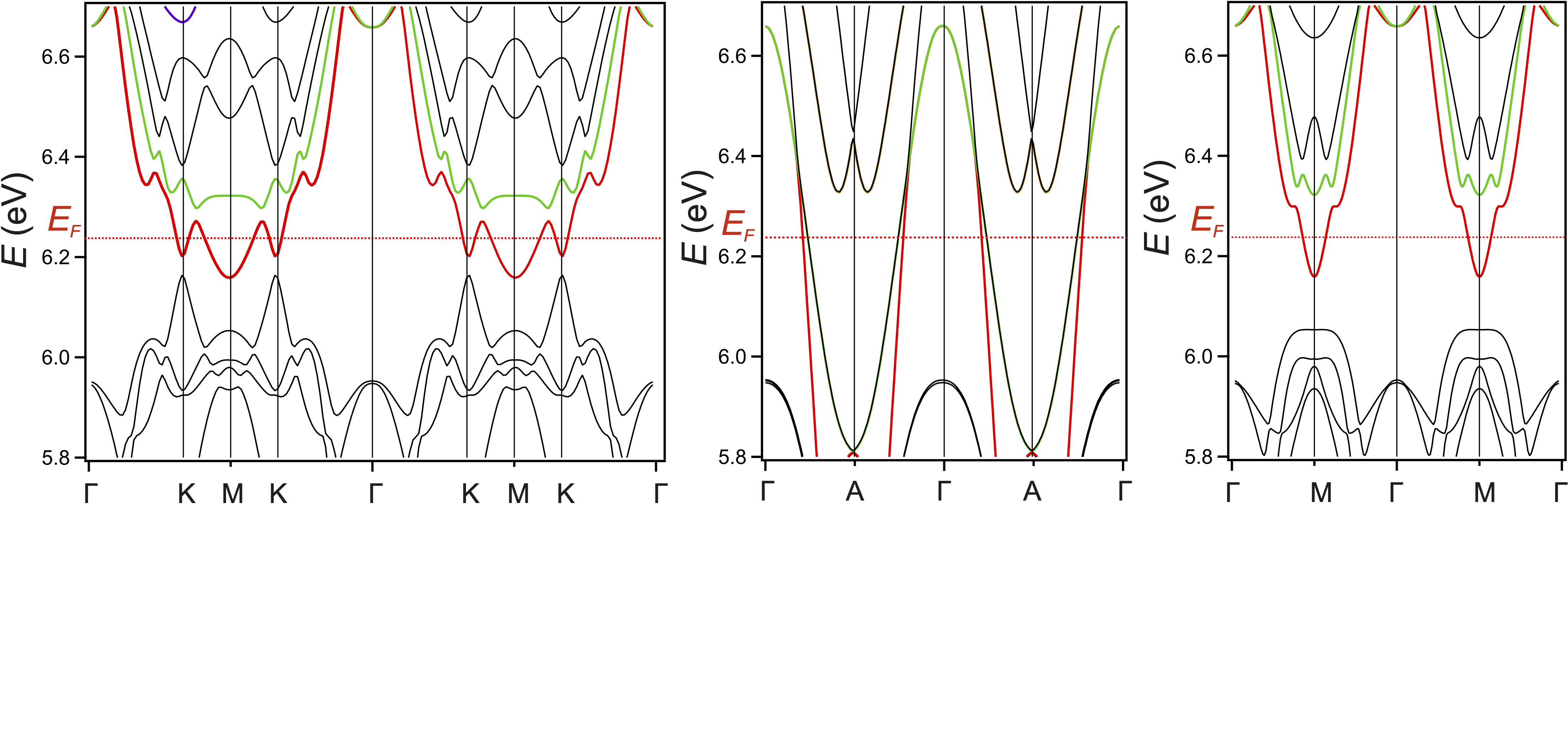}
	\caption{\label{fig:E_Band}
		Calculated electronic band structures and the Fermi level $E_{F}$ of the stoichiometric Mn$_{3}$Sn, in which both SOC and the room-temperature $\mathbf{k}$ = 0 magnetic order were included in the calculations. The relevant electronic bands near $E_{F}$ are highlighted in red and green colour, respectively. 
	}
\end{figure}

\begin{figure}[htbp]
	\centering
	\includegraphics[width=8.5cm]{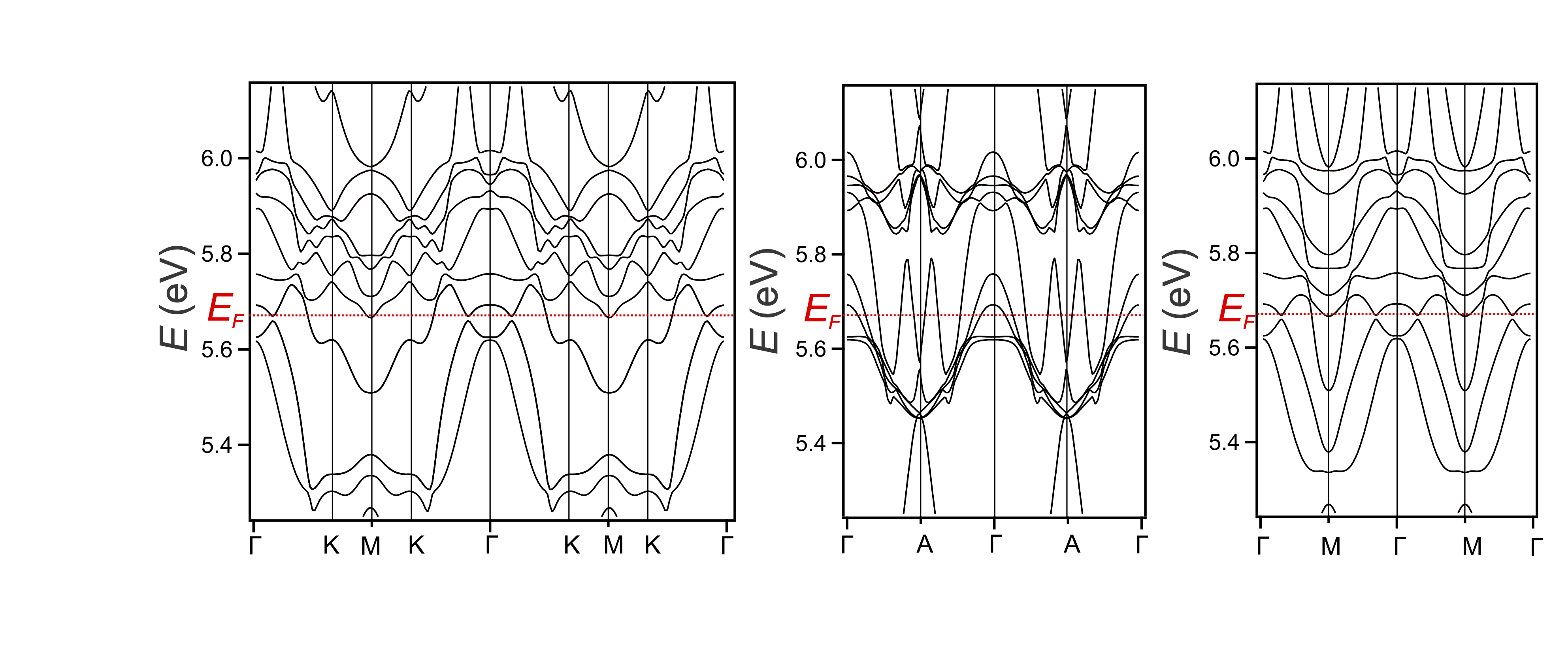}
	\caption{\label{fig:E_Band_no_mag}
		Calculated electronic band structures and the Fermi level $E_{F}$ of the stoichiometric Mn$_{3}$Sn, in which only SOC was included in the calculations, but not the room-temperature $\mathbf{k}$ = 0 magnetic order.		
	}
\end{figure}

\begin{figure}[htbp]
	\centering
	\includegraphics[width=8.5cm]{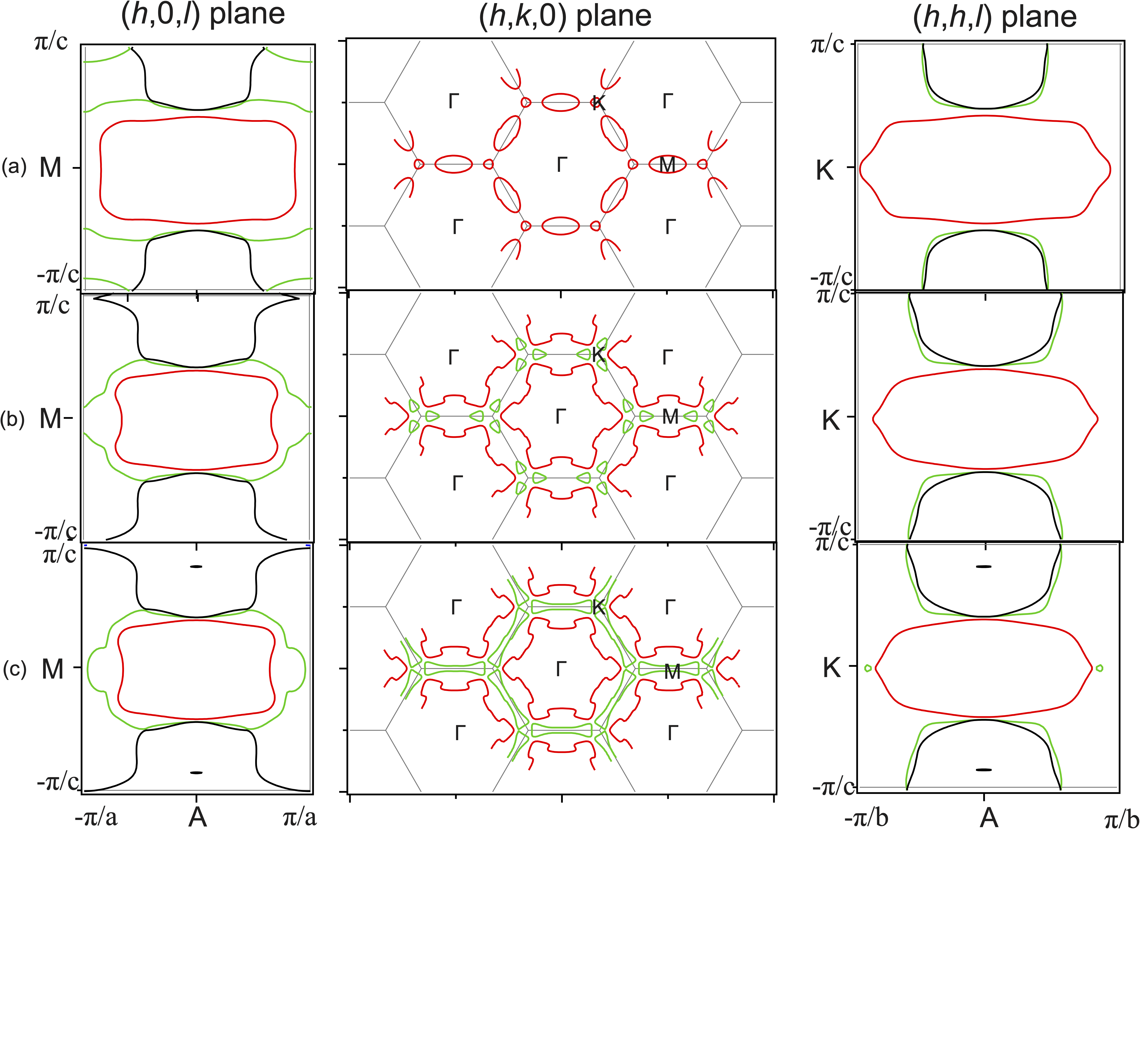}
	\caption{\label{fig:E_Band2}
		Plotted Fermi surfaces in the ($h$,0,$l$), ($h$,$k$,0) and ($h$,$h$,$l$) reciprocal-space planes for the weakly doped case as in our studied sample (i.e., with $E_{F}$+81 meV), in which no Fermi surface nesting conditions could be found in the plotted reciprocal-space planes. The relevant electronic bands near $E_{F}$ (see Fig.\ref{fig:E_Band}) are shown in red and blue colour, respectively. 
	}
\end{figure}

\begin{figure}[htbp]
	\centering
	\includegraphics[width=7cm]{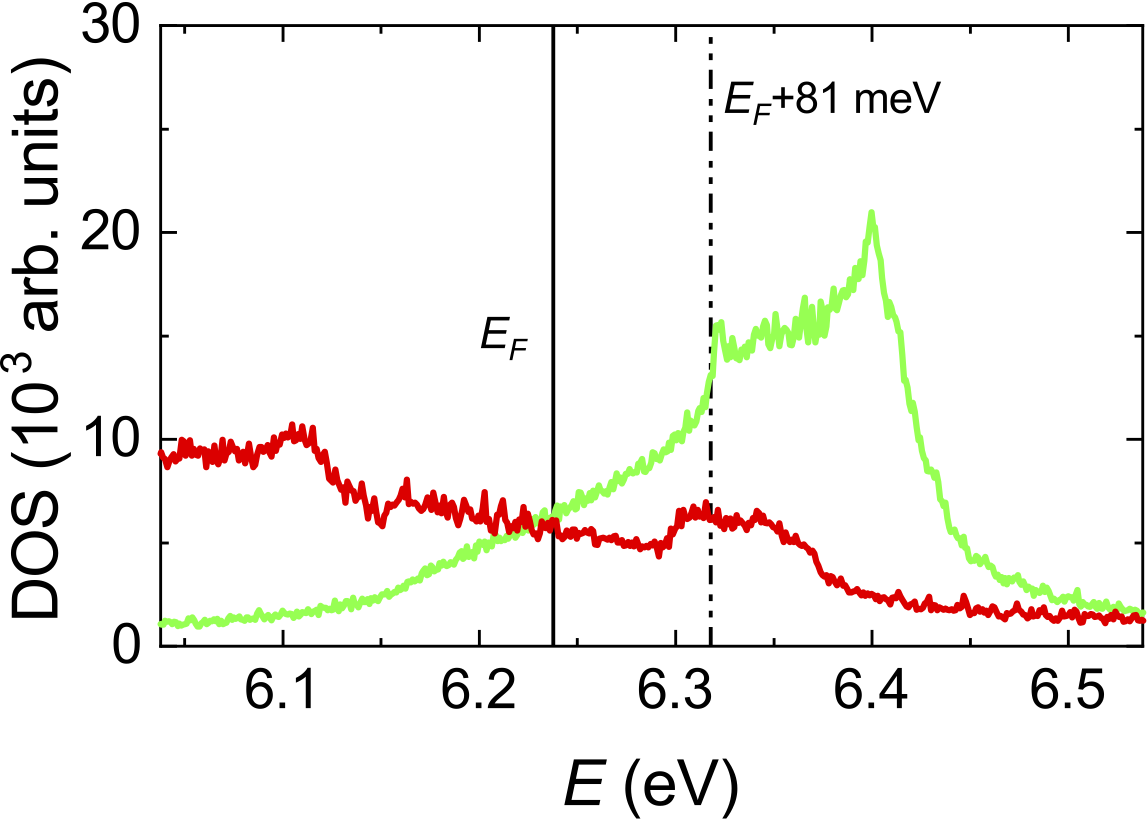}
	\caption{\label{fig:DOS}
		Calculated density of states (DOS) of the relevant electronic bands near $E_{F}$ (see Fig.\ref{fig:E_Band}), in which a sharp local maximum in DOS in the green-coloured electronic band, due to the presence of the \textit{K}-\textit{M}-\textit{K} flat band, can be clearly seen at the shifted Fermi level $E_{F}$+81 meV, but, no such features exist in the red-coloured electronic band.
	}
\end{figure}

The electronic band structures of Mn$_{3}$Sn were comprehensively studied based on the first-principles DFT calculations. The calculated electronic band structures and the Fermi level $E_{F}$ of the stoichiometric Mn$_{3}$Sn, in which both spin-orbit coupling (SOC) and the room-temperature $\mathbf{k}$ = 0 magnetic order were included in the calculations, is shown in Fig.\ref{fig:E_Band}. The room-temperature $\mathbf{k}$ = 0 magnetic structure is based on our own single-crystal neutron diffraction refinements in Tab.\ref{tab:mag_300K}. For a comparison, the calculated electronic band structures, in which only SOC was included in the calculations but not the room-temperature $\mathbf{k}$ = 0 magnetic order, is shown in Fig.\ref{fig:E_Band_no_mag}. A significant impact of magnetism on the dispersion and degeneracy of band structure near $E_{F}$ can be clearly seen in this comparison. Furthermore, as shown in Fig.\ref{fig:E_Band2}, the Fermi surfaces in the ($h$,0,$l$), ($h$,$k$,0) and ($h$,$h$,$l$) reciprocal-space planes for the weakly doped case as in our studied sample, in which $E_{F}$ is shifted up by 81 meV, are also plotted for a comparison to Fig.5(d-f) in the main text. No Fermi surface nesting conditions could be realised in these reciprocal-space planes. The calculated density of states (DOS) of the relevant electronic bands near $E_{F}$ are plotted in Fig.\ref{fig:DOS}, in which a sharp local maximum in DOS due to the presence of the \textit{K}-\textit{M}-\textit{K} flat band can be clearly seen at the shifted Fermi level $E_{F}$+81 meV. 

\cleardoublepage

\bibliography{Mn3Sn_combined_new.bib}

\end{document}